\documentclass[12pt,preprint]{aastex}
\usepackage{rotating}
\usepackage{graphicx}
\newcommand {\mdot}{m\kern-0.1667em\bdot}

\shorttitle{The SDSS/XMM-Newton Quasar Survey}
\shortauthors{M. Young et al.}

\received{12.11.2008}
\begin{document}

\title{The Fifth Data Release Sloan Digital Sky Survey/XMM-Newton Quasar Survey}

\author{
M. Young\altaffilmark{1,2}, M. Elvis\altaffilmark{1}, 
G. Risaliti\altaffilmark{1,3}
} 
\email{myoung@cfa.harvard.edu}

\altaffiltext{1}{Harvard-Smithsonian Center for Astrophysics, 60 Garden St. 
Cambridge, MA 02138 USA}
\altaffiltext{2}{Boston University, Astronomy Department, 725 Commonwealth Ave., 
Boston, MA 02215}
\altaffiltext{3}{INAF - Osservatorio di Arcetri, L.go E. Fermi 5,
Firenze, Italy}
\begin{abstract}
We present a catalog of 792 DR5 SDSS quasars with optical spectra that have been observed serendipitously 
in the X-rays with the \emph{XMM-Newton}.  These quasars cover a redshift range of z = 0.11 - 5.41 and a 
magnitude range of $i$ = 15.3-20.7.  Substantial numbers of radio-loud (70) and broad absorption line (51) 
quasars exist within this sample.  Significant X-ray detections at $\geq2\sigma$ account for 87\% of the 
sample (685 quasars), and 473 quasars are detected at $\geq$ 6$\sigma$, sufficient to allow X-ray spectral 
fits.  For detected sources, $\sim$ 60\% have X-ray fluxes between F$_{2-10keV}$ =  1 - 10 x 10$^{-14}$ ergs 
cm$^{-2}$ s$^{-1}$.  We fit a single power-law, a fixed power-law with intrinsic absorption left free to vary, 
and an absorbed power-law model to all quasars with X-ray S/N $\geq$ 6, resulting in a weighted mean photon 
index $\Gamma$ = 1.91$\pm$0.08, with an intrinsic dispersion $\sigma_{\Gamma}$ = 0.38.  For the 55 sources 
(11.6\%) that prefer intrinsic absorption, we find a weighted mean N$_H$ = 1.5$\pm$0.3 x 10$^{21}$ 
cm$^{-2}$.  We find that $\Gamma$ correlates significantly with optical color, $\Delta$($g - i$), the 
optical-to-X-ray spectral index ($\alpha_{ox}$) and the X-ray luminosity.  While the first two correlations 
can be explained as artefacts of undetected intrinsic absorption, the correlation between $\Gamma$ and X-ray 
luminosity appears to be a real physical correlation, indicating a pivot in the X-ray slope.
\end{abstract}


\section{Introduction}

Catalogs are indispensable in performing statistical studies of quasar properties.  The known 
correlations between optical and X-ray properties, discussed further below, imply a connection 
between the accretion disk posited to feed the central black hole \citep{SS73} and the hot 
Compton-scattering corona posited to lie in some unknown geometry around the disk \citep{HM91}.  
While quasars were first discovered in early radio surveys (e.g. 3C, 3CR, PKS, 4C, AO), most 
quasar surveys since then have been conducted in the optical/UV.  Early quasar selection 
techniques included UV-excess selection and emission line searches using slitless spectroscopy.  

UV-excess selection 
utilizes the Big Blue Bump that dominates the optical/UV spectrum to distinguish quasars from 
stars.  The Big Blue Bump is normally attributed to a multi-temperature accretion disk 
\citep{Shields78,MS82}.  UV-excess based surveys include the \citet{Braccesi70} catalog, which 
contains 175 objects with U - B  $<$ -0.42, and the Palomar-Green (PG) Bright Quasar Survey 
\citep[BQS,][]{SG83}, a sample of 92 objects selected with U - B $<$ -0.44.  
Slitless spectroscopy, with prisms or 
grisms, obtains a large number of low-resolution spectra for a single field.  For example, the 
objective prism on the Curtis Schmidt telescope at CTIO found 174 confirmed quasars, ranging in 
redshift from z = 0.1 to z = 3.3 \citep{OS80,Osmer81}.  Slitless spectroscopy often incorporated 
UV-excess selection as well; the primary example of this technique is the Markarian survey 
\citep{Markarian67,Markarian81}, which searched for galaxies with unusually blue continua using 
a grism.  The Large Bright Quasar Survey (LBQS; Hewett et al. 1995) used a combination 
of color selection and the presence of emission features on objective prism plates to obtain a homogenous 
sample of 1055 quasars spanning a wide redshift range (0.2 $\leq$ z $\leq$ 3.4).  

However, both of these techniques suffer from serious biases.  While slitless spectroscopy has a high 
selection efficiency, even at higher redshifts, it is biased against quasars with weak emission lines, 
and cannot reach as faint a flux limit.  The UV-excess selection method is biased against "red" sources, 
where red colors may be due to high-redshift (the Ly$\alpha$ line enters the spectrum at z $\sim$ 2), 
dust-reddening, significant host galaxy contribution, or intrinsically red emission mechanisms.  With 
the advent of the `UK Schmidt Survey' \citep{Warren91}, the Two-Degree Field \citep[2DF;][]{Croom01} 
and the Sloan Digital Sky Survey (SDSS; York et al. 2000) quasar catalogs, multicolor selection techniques 
that used up to 5 photometric bands were introduced that could select red quasars in addition to blue ones, 
provided the sources were within the survey flux limit.  The 5th Data Release (DR5) SDSS quasar 
catalog has surpassed all previous optical surveys by providing high quality photometry and spectroscopy 
for 77,429 quasars \citep{Schneider07} spanning redshifts from z = 0.08 to z = 5.41.  

As quasars emit strongly over eight decades of the spectrum \citep[e.g.][]{Elvis94}, multiwavelength surveys 
are necessary to relate the optical/UV accretion emission to other components, notably the non-thermal 
emission seen in the X-rays.  However, X-ray spectra are time-consuming and expensive to obtain for large 
samples.  For this reason, previous studies of optical and X-ray correlations consist largely of two types:  
(1) Small samples (N $\sim$ 20-50) with detailed X-ray spectral analysis have been compiled by observing 
sub-samples of optical surveys with X-ray telescopes \citep[e.g.][]{Laor97,Elvis94,Piconcelli05,Shemmer06,
Shemmer08}.  The \citet{Akylas04} \emph{XMM-Newton}/2dF survey is larger, with 96 2QZ quasars 
observed in wide field (2.5 deg$^2$), shallow (2-10 ks per pointing, \emph{f}(0.5-8 keV) $\sim$ 10$^{-14}$ 
erg cm$^{-2}$ s$^{-1}$) \emph{XMM} observations.
(2) Still larger samples (N $\sim$ 200-300) with only X-ray fluxes have been compiled for statistical 
investigations.  These larger studies have to assume an X-ray spectral slope \citep[e.g.][]{Vignali03,
Strateva05,Steffen06}.  

The ever-expanding archive of X-ray observations now provides a less costly and time-consuming method of 
obtaining X-ray spectra for large numbers of optically-selected quasars.  \citet{Kelly07} cross-correlated 
the SDSS DR3 quasar catalog with archival \emph{Chandra} observations \citep{Weisskopf99} to obtain 174 
quasars, 44 of which have sufficient counts to fit an absorbed power-law with $\Gamma$ and N$_H$ as free 
parameters.  This sample was extended to a total of 318 quasars \citep{Kelly08} by adding 149 RQ objects 
from an SDSS-ROSAT cross-correlation \citep{Strateva05}.  As only single power-law models were fit to the 
ROSAT spectra with sufficient counts, 153 of 318 sources have X-ray spectra slopes, but no fits for intrinsic 
absorption were made.  Archival \emph{Chandra} observations have also been used to identify an X-ray-selected 
AGN population in the \emph{Chandra} Multiwavelength Project \citep{Green03}.  In addition, a large sample 
(N = 1135) of optically selected SDSS quasars with photometric redshifts are detected in the X-rays in 
\emph{Chandra} fields \citep{Green08}, of which 156 have sufficient counts ($> 200$) to fit an absorbed 
power-law with $\Gamma$ and N$_H$ as free parameters.    

\emph{XMM-Newton} is a good choice for cross-correlating with optical catalogs due to its large field of 
view and large effective area.  
Images are made with the three European Photon Imaging Cameras (EPIC): MOS-1 and MOS-2 \citep{Turner01}, 
each of which has a 33' x 33' field of view, and the pn camera \citep{Strueder01}, which has a 27.5' x 27.5' field of 
view.\footnote{http://imagine.gsfc.nasa.gov/docs/sats\_n\_data/missions/xmm.html}.  \emph{XMM}'s large 
effective area (922 cm$^2$ for MOS and 1,227 cm$^2$ for PN at 1 keV)$^4$ results in higher signal-to-noise 
X-ray spectra than with \emph{Chandra} for bright, non-background-limited sources\footnote{http://heasarc.
nasa.gov/docs/xmm/uhb/node86.html}.  The EPIC CCDs have good spectral resolution (E/$\Delta$E $\sim$ 
20 - 50 for both MOS and PN) over the 0.5 - 10 keV band.  

Archival \emph{XMM-Newton} observations overlapped with $\sim$1\% of the SDSS DR5 coverage as of Feb. 2007.  
The SDSS and \emph{XMM-Newton} archives are well-matched in sensitivity, as discussed in \citet{Young08}.  
The 5th Data Release \citep[DR5,][]{AM07} of the SDSS covers a spectroscopic area of 5740 deg$^2$, 
and contains 90,611 quasars.  The SDSS photometry in the $u$, $g$, $r$, $i$ and $z$ bands covers 
3,250 - 10,000 $\mbox{\AA}$, while the spectroscopy covers 3,800 - 9,200 $\mbox{\AA}$ with a spectral 
resolving power of $\sim$ 2,000.  The SDSS is $\sim$95\% complete for point sources to a limiting 
magnitude $i = 19.1$, corrected for Galactic reddening \citep{Richards02,VdB05}.

A preliminary cross-correlation of DR1 SDSS quasars with the \emph{XMM-Newton} public archive yielded 55 
objects with exposure times greater than 20 ks \citep{Risaliti05}.  Of these, 35 
yielded good X-ray spectra.  \citet{Risaliti05} estimated that a cross-correlation of a final 
SDSS data release with the ever-growing \emph{XMM-Newton} archive would yield $\sim$ 1000 quasars, 
of which $\sim$80\% would have good X-ray spectra.

In this paper, we cross-correlate the SDSS DR5 quasar catalog with archival \emph{XMM-Newton} observations 
to obtain a large (N$\sim$800) sample of quasars with X-ray detections, almost 500 of which have good optical 
and X-ray spectral data.  Below, we outline two immediate goals for the \emph{SDSS/XMM-Newton} Quasar Survey: 
(1) to conduct large, statistical studies to understand the physical basis behind optical/X-ray correlations 
and (2) to investigate interesting sub-populations of quasars.

\subsection{\it Optical/X-ray Correlations}

The relations between optical and X-ray continuum and spectral properties promise to reveal clues 
about the disk-corona structure of quasars.  The large sample provided by the \emph{SDSS/XMM-Newton} 
Quasar Survey allow two correlations to be investigated.  
The first of these is the controversial $\alpha_{ox}-l_{2500\AA}$ relation: many studies have found that 
the spectral index from 2500 $\mbox{\AA}$ to 2 keV, defined as $\alpha_{ox}$ = log(L$_{2keV}$/
L$_{2500}$)/log($\nu_{2keV}$/$\nu_{2500}$), anti-correlates with the log of the monochromatic luminosity 
at 2500 $\mbox{\AA}$, $l_{2500\AA}$ \citep{Tananbaum79,Zamorani81,AT82,KC85,Tananbaum86, 
AM87, Wilkes94, PIF94, AWM95,Vignali03, Strateva05, Shen06, Steffen06, Just07}.  
However, some studies \citep{Bechtold03, Kelly07} find the primary relation to be between $\alpha_{ox}$ 
and redshift, while other studies \citep{Yuan98, Tang07} find that the correlation may be induced by 
selection effects.  

No physical basis for the $\alpha_{ox}-l_{2500\AA}$ relation has yet been proposed, and the relation itself 
provides little guidance.  In part, this is because previous studies have largely used the traditional, 
observationally convenient, but physically arbitrary, endpoints of 2500~$\mbox{\AA}$ and 2 keV.  They also 
assume an X-ray photon index ($\Gamma \sim$ 2, where $\Gamma$ = -$\alpha$ + 1 for F$_{\nu} \propto \nu^{\alpha}$) 
to obtain the X-ray flux at 2 keV.  A systematic study with measured optical and X-ray spectra would enable an 
investigation of the relation at different frequencies than those traditionally used, hopefully revealing clues 
about the relation's physical underpinnings (Young et al. 2009b, in prep.)\nocite{Young09b}.  

The second correlation is the positive relation between $\Gamma$ and the 
normalized accretion rate, L/L$_{Edd}$ \citep{Shemmer06, Shemmer08}.  While early studies focused on the 
relation between $\Gamma$ and full-width half-maximum of the H$\beta$ line, FWHM(H$\beta$) \citep{Boller96, 
Laor97, Brandt97}, Shemmer et al. (2006, 2008)\nocite{Shemmer06, Shemmer08} have broken the degeneracy 
between FWHM(H$\beta$) and L/L$_{Edd}$ by including highly luminous sources in order to show that $\Gamma$ 
depends primarily on L/L$_{Edd}$.  There have been suggestions that this dependence may be due to the disk 
emitting more and softer photons as accretion rates increase, leading to more efficient Compton 
cooling in the corona \citep{Laor00,Kawaguchi01}.  

Until now, studies of the $\Gamma$-L/L$_{Edd}$ relation have consisted of small samples (N $\sim$ 40).  
The \emph{SDSS/XMM-Newton} quasar survey can increase this sample size by an order of magnitude, leading to 
a better defined relation (Risaliti et al. 2009, in prep.)\nocite{Risaliti09}.

\subsection{\it Quasar Sub-Populations}

Interesting quasar sub-populations can be readily investigated due to the large number of sources 
in the SDSS/\emph{XMM-Newton} Quasar Survey.  For example, red quasars make up 6\% of the SDSS sample 
\citep{Richards03}.  Their steep optical slopes have been attributed to dust-reddening \citep{Richards03, 
Hopkins04}, though observations of individual objects suggest that some slopes may be intrinsically 
steep \citep{Risaliti03,Hall06}.  In \citet{Young08}, we studied 17 quasars with extreme red colors 
($g - r > 0.5$) and moderate redshifts (1 $<$ z $<$ 2).  By using X-ray observations in conjunction 
with optical spectra, we constrained the amount of intrinsic absorption in each source, thereby allowing 
the separation of intrinsically red from dust-reddened optical continua.  We find that almost half (7 of 17) 
of the quasars can be classified as probable `intrinsically red' objects.  These quasars have unusually 
broad MgII emission lines ($<$FWHM$>$=10,500 km s$^{-1}$), flat but unabsorbed X-ray spectra ($<\Gamma>$ 
= 1.66$\pm$0.08), and low accretion rates ($\dot{M}/\dot{M_{Edd}} \sim$ 0.01).  

Other interesting sub-populations for future investigations include broad absorption line (BAL), Type 2, and 
radio-loud (RL) quasars.

In this paper, we describe the SDSS quasar selection and the method with which we match sources to 
XMM-Newton observations in \S2.  X-ray data reduction is described in \S3 and the resulting sample and 
correlations are discussed in \S4.  We assume a standard cosmology throughout the paper, where H$_0$ = 
70 km s$^{-1}$ Mpc$^{-1}$, $\Omega_M = 0.3$, and $\Omega_\Lambda = 0.7$ \citep{Spergel03}. 

\section{Data}

\subsection{\it SDSS Quasar Selection}

As the DR5 quasar catalog \citep{Schneider07} was not yet available at the time of selection, 
we selected quasars with optical spectra directly from the DR5 SDSS database\footnote{http://
www.sdss.org/dr5/access/index.html} by choosing SpecClass = 3 (QSO) or 4 (QSO with z $>$ 2.3, 
whose redshift has been confirmed using a Ly$\alpha$ estimator).  These quasars were selected 
for spectroscopic follow-up by the SDSS primarily due to their photometric colors, although 
some quasars were selected because they have a match in the FIRST survey \citep{White97}.  
No X-ray selection is involved.  Target selection efficiency, i.e. the percentage of sources 
spectroscopically observed that are confirmed as quasars, is 66\% \citep{Richards02, VdB05}.  
The selection method is described in detail by \citet{Richards02}, and is summarized briefly below.

The vast majority ($\sim$95\%) of SDSS quasar candidates are detected using a multicolor 
selection technique.  Quasar candidates are defined to be any object at least 4$\sigma$ away 
from the stellar locus, which is defined in the ($u - g$,$g - r$,$r - i$,$i - z$) color 
space.  In addition, special color-color regions are defined to specifically include or exclude 
quasar candidates.  Inclusion regions include quasar candidates from 2.5 $<$ z $<$ 3, even if their 
colors cross the stellar locus.  Exclusion regions prevent contamination due to white dwarfs, 
A stars and M star-white dwarf pairs.  

For both radio and color-selected quasar candidates, magnitude limits are applied.  Quasar candidates 
brighter than $i$ = 15 are rejected for spectroscopic follow-up because bright sources can contaminate 
the spectra of objects in adjacent fibers in the SDSS spectograph.  Radio and low-redshift color-selected 
candidates fainter than $i$ = 19.1, which have a high number density on the sky, are also rejected due 
to the limited number of optical fibers for follow-up spectroscopy.  Since high redshift (z $\gtrsim$ 3) 
quasar candidates have a lower surface density on the sky, a fainter cut-off magnitude $i$ = 20.2 is 
applied to these objects.  

SDSS selected a small number of quasar candidates ($\sim$5\%) by matching point sources 
with the position of a radio detection from the FIRST survey to within 2''.  While the 
DR5 quasar catalog \citep{Schneider07} includes sources selected by matching to \emph{ROSAT} 
detections, the initial SDSS quasar selection outlined in \citet{Richards02} does not use 
X-ray detection as a criterion.  

While SDSS radio selection requires that a source be point-like, their color selection includes extended 
sources as well, in order to include low-redshift AGN, such as Seyfert galaxies.  However, in addition 
to having colors distinct from the stellar locus, extended sources must also have colors distinct from 
the main galaxy distribution.  (The main galaxy distribution overlaps the stellar locus; however, a galaxy 
can be a clear outlier from the stellar locus both due to the shape of the stellar locus and because the 
stellar locus is determined primarily from F and M stars that dominate the Galactic stellar density at 
high latitudes.)  Simple color cuts are applied to distinguish extended sources from galaxies, rather 
than an additional multicolor selection.  

The DR5 quasar catalog \citep{Schneider07} additionally requires: 
luminosities brighter than M$_i$ = -22.0; at least one emission line with a FWHM greater than 1000 
km s$^{-1}$ or interesting/complex absorption features; magnitudes fainter than i = 15.0; and highly 
reliable redshifts.  This results in a catalog of 77,429 quasars.  We used only those quasars that 
are also in the DR5 Quasar Catalog for further analysis.  Of the 92 sources we reject from our sample, 
$\sim$40\% are Type 2 quasars.  These will be the subject of a later study.

We matched The selected SDSS quasars with the FIRST survey~\citep{White97}\footnote{http://sundog.stsci.edu/} 
using a search radius of 3'' in order to calculate their radio-loudness (R$_L$ = F$_{5GHz}$ / F$_{4400}$, 
Kellerman et al. 1989)\nocite{Kell89}).  A source is taken to be radio loud (RL) if R$_L >$ 10.  A power-law 
is interpolated between the optical magnitudes to get F$_\lambda$(4400$\mbox{\AA}$), and the 1.4~GHz radio 
flux is obtained from FIRST survey detections, which are extrapolated to 5~GHz using a radio power-law $\alpha_R$ 
= -0.8.  All the quasars lie in the area covered by the FIRST survey, so if there is no detection, we use 
the 5$\sigma$ upper-limit on the 1.4 GHz radio flux to extrapolate to 5~GHz.  For 40 quasars, the upper-limit 
is too high to determine whether the source is radio-loud.  Of the remaining quasars, 70 (9.3\%) are 
radio-loud and 682 (90.7\%) are radio-quiet, a typical ratio (e.g. Kembhavi \& Narlikar 1999, pp. 256-263)\nocite{KN99}.  

The SDSS color-color selection is effective at finding a variety of Broad Absorption Line (BAL) quasars 
\citep{Schneider07}.  The official DR5 BAL catalog \citep{Gibson09} had not yet been published during the 
writing of this paper, so we used an incomplete list of $\sim$ 4200 BALs in the DR5 Quasar Catalog 
\citep{Shen08a}, selected using traditional criteria \citep{Weymann91}.  Using \citet{Shen08a}, we find 52 
BALs in the \emph{SDSS-XMM} Quasar Survey. Of these, 15 BALs have high enough X-ray signal-to-noise to obtain 
X-ray spectra (see \S3).  Since BALs are normally identified by CIV absorption features, which are visible 
only above a redshift z = 1.6 in SDSS spectra, BALs are unlikely to be identified in $\sim$56\% of the 
\emph{SDSS-XMM} quasars.  BALs make up 14.5\% of the \emph{SDSS/XMM} quasars with z $>$ 1.6, 
which is in line with the parent population in \citet{Shen08a}.  

In this paper, we measure the optical color of a quasar by its relative color \citep{Richards03}.  Relative 
colors compare a quasar's measured colors with the median colors in its redshift bin, where redshift bin 
sizes are 0.1 in redshift, so that $\Delta$($g - i$) = ($g - i$) - $<$($g - i$)$>_z$.  The use of relative colors 
corrects for the effect of typical emission lines on the photometry in a particular band.  The relative 
($g - i$) colors of the SDSS quasars match a Gaussian distribution on the blue side but require the addition of a 
tail on the red side [see Fig. 3 in \citet{Richards03}].  

\subsection{\it Matching with the XMM-Newton Archive}

We matched the SDSS quasars with the \emph{XMM-Newton} archive from February 2007, choosing 
only those quasars that fell within 14' of XMM observation field centers (typically 2-3 quasars 
per field).  As part of the extraction process, described below, a 
source region is defined around each set of SDSS quasar coordinates.  Depending on the S/N of the 
source, the extraction radius can range from 10 - 85", with a typical radius of 19".  Low S/N 
objects are extracted with smaller radii to minimize the effect of high background levels, while 
larger radii for high S/N objects allow for an increased encircled energy fraction in the presence 
of relatively low background levels.  The large extraction radii take into account the \emph{XMM} 
point spread function (PSF), which is characterized by the radius at which 90\% of  the total energy 
is encircled.  This radius increases from 48" (MOS) and 51.5" (PN) at 0' off-axis angle to 52.5" (MOS) 
and 66" (PN) at 12' off-axis angle\footnote{http://xmm.vilspa.esa.es/external/xmm\_user\_support/documentation
/uhb/node17.html}.  The extraction radii also take positional accuracy into account: SDSS positional 
errors are negligible (0.1'' at the survey limit of r=22 for typical seeing, Pier et al. 2003\nocite{Pier03}), 
while the \emph{XMM} positional accuracy is 3" at 3$\sigma$ for offset angles 0 $< \theta <$ 10 arcmin and 
6" for 5 $< \theta <$ 10 arcmin \citep{Pierre07}.  

Multiple X-ray observations exist for 265 sources.  In these cases, all observations were retrieved 
and reduced, but only the observation with the longest exposure time was used for further analysis.  
To avoid biases, we did not select the highest S/N observations, although in $\sim$80\% cases, the two 
selections would be effectively the same.  

The chance of including an unrelated random source within the extraction region is small but non-zero.  
Within a $\sim$14' radius field of view, and using the average extraction radius, 19'', there are 1954 
``beams" in an observation.  Since there are $\sim$70 sources in a typical \emph{XMM} observation 
\citep[][which has an exposure time distribution similar to that in this paper]{Watson08}, this results 
in a 3.6\% chance of extracting a random source rather than the SDSS-selected quasar.  Since a random 
source is likely to be faint in the X-rays, any contamination is only significant for sources 
under 100 net counts.  We have 792 unique sources, of which 390 have less than 100 net counts, so $\sim$14 
(2\% of the \emph{SDSS/XMM} quasar sample) have significant contamination from an unrelated `interloper' source.

\subsection{\it X-ray Data Reduction} 

The 582 X-ray observations were processed using the \emph{XMM-Newton} Science Analysis System, SAS 
v7.02\footnote{http://xmm.esac.esa.int/sas}.  We reprocessed the events to ensure 
that each observation has the same, up-to-date calibration, and then filtered the 
observations to remove time intervals of flaring high-energy background events using 
the standard cut-off of 0.35 cts s$^{-1}$ for the MOS cameras and 1.0 ct s$^{-1}$ 
for the PN camera\footnote{http://xmm.vilspa.esa.es/sas/7.1.0/documentation/threads/}.  
Source and background regions were defined in a semi-automatic process.  
The SAS task \emph{eregionanalyse} was used to optimize the source extraction radius for 
signal-to-noise.  Most radii include at least 80\% of the source counts.  
Background regions were defined by eye, avoiding obvious X-ray sources and chip edges.  
These regions were typically a circle of radius 2000-2500 pixels (100-125"), selected 
to lie at the same off-axis angle as the source and as close to the source as possible without 
overlapping the source extraction region.  Once source and background regions were defined 
for every SDSS quasar in an observation, spectra were extracted for a total of 1380 non-unique 
quasars in 582 observations.  

To check for biases in the data reduction, Figures 1$a - b$ show the net counts and the X-ray 
photon index ($\Gamma$) plotted against the extraction radius.  Figures 1$c - d$ plot the 
off-axis angle and the X-ray S/N against $\Gamma$.  In Fig. 1$a$, the net counts are expected 
to correlate with the extraction radius, since the extraction radius will increase to larger 
encircled energy fraction for sources that stand higher above the background.  The lack of 
correlations in Fig. 1$b$-$c$ show that the encircled energy correction takes the extraction 
radius and off-axis angle into account correctly.  Fig 1$d$ shows that objects with flat $\Gamma$ 
are primarily found among low S/N objects, where absorption may be undetected in a spectral fit.  

Where possible, observations were processed for all three XMM EPIC CCDs.  In $\sim$40\% of the observations, 
a source lies in a bad region in one or two of the three cameras, either in a strip between two chips 
or, because the MOS and PN cameras have different shapes, outside the field of view in one of the 
cameras.  In these cases, we use the remaining images from the other cameras for analysis.  

Table~\ref{table:Xray obs} contains observational data for each quasar in the DR5 \emph{SDSS/XMM-
Newton} Survey: the SDSS name, XMM observation ID, redshift, Galactic column density in the 
direction of the source, X-ray signal-to-noise, observation exposure time, off-axis angle, net 
source counts, background counts (from a background region that is scaled to the area of the 
source region), and two flags indicating if a source is RL or BAL.  

Figure~\ref{fig:characteristics} summarizes the survey characteristics.   The X-ray exposure times 
(Fig. 2$a$) range from 1.6 to 294 kiloseconds, though the majority of observations lie between 20 
and 100 kiloseconds.  This range in exposure times results in a wide range in sensitivity.  While 
most sources have low signal-to-noise, a significant fraction have S/N $\geq$ 10, where more complex 
models can be fit (Fig. 2$b$).  The detection fraction is $\gtrsim$ 80\% until z $>$ 3.5 (Fig. 2$c$), 
but spectral coverage drops off fairly quickly for sources with z $>$ 2 (Fig. 2$d$).  

\section{X-ray Analysis: Spectral Fits}

We made fits to the extracted spectra using the \emph{Sherpa} package\footnote{http://cxc.harvard.edu/
sherpa/threads/index.html} within CIAO\footnote{http://cxc.harvard.edu/ciao/}.  For each source, the 
available MOS+PN spectra were fit simultaneously over the 0.5 - 10 keV band.  The observations were fit 
according to their S/N, with more complicated models being applied as S/N increased.  All the models 
included local absorption fixed to the Galactic hydrogen column density (N$_{H,gal}$) at 
each source location.  Values for N$_{H,gal}$ were taken from the N$_H$ tool available at 
WebPIMMS\footnote{http://heasarc.gsfc.nasa.gov/Tools/w3pimms.html}, which is based on the 21 cm HI 
compilation of Dickey \& Lockman (1990) and Kalberla et al. (2005)\nocite{DL90,Kalberla05}.

For the 319 low S/N sources (S/N $\leq$ 6), we fix a power-law to the weighted mean obtained for the 
high S/N (S/N $\geq$ 6) quasars in the sample, $\Gamma \sim$ 1.9 (\S4.1), and allow only the normalization 
to vary in order to obtain the flux.  For the 101 sources with S/N $<$ 2, we obtain a 90\% upper-limit 
to the flux.  We use the Cash (1979)\nocite{Cash79} statistic, which gives more reliable results for 
low-count sources, to fit sources with S/N $\leq$ 6.  As the background is not subtracted and is instead fit 
simultaneously with the source, we apply a background model with three components, as described in 
\citet{Lumb02} and in the XMM Users’ Handbook\footnote{http://xmm.vilspa.esa.es/external/xmm\_user\_
support/documentation}: a power-law (for the extragalactic X-ray spectrum), a broken power-law (for 
the quiescent soft proton spectrum), and two spectral lines (for cosmic-ray interactions 
with the detector).  All parameters except for the normalization of the spectral lines were fixed.  
For 13 undetected sources, and 5 detected sources, this fit results in bad or null flux values.  
In these cases, we flag the source as undetected (flag = -1 in column (9) of Table \ref{table:specdata}) 
and we list the $\alpha_{ox}$ values as 9.99 (column (8) of same table).  

For the 473 sources with enough S/N to fit spectral parameters (S/N $\geq$ 6), we use the $\chi^2$ 
statistic to fit three models: a single power-law (SPL) with no intrinisc absorption, a fixed 
power-law (FPL) where intrinsic absorption is left free to vary, and an intrinsically absorbed 
power-law (APL).  The F-test\footnote{http://cxc.harvard.edu/ciao/ahelp/ftest.html}, which measures 
the significance of the change in $\chi^2$ as components are added to a model, is used to determine 
whether the data prefer the APL model.  To compare the SPL and FPL models, we simply compare the 
respective $\chi^2_\nu$ values, since both models have the same number of parameters.  Therefore, 
the best-fit model is the one preferred by the F-test that also has the lowest $\chi^2_\nu$ value.  

For sources with an unacceptable $\chi^2_\nu$ 
values for all three models, we by default make the SPL model the best-fit, but we also list the 
APL 90\% upper-limit on intrinsic absorption.  We plot the reduced $\chi^2$ distribution in Figure 
\ref{fig:chi2}.  A later paper will look in more detail at sources with bad fits ($\chi_\nu^2 > 1.2$).

Results for the best-fit models are listed in 
Table~\ref{table:specdata}, including observed-frame and rest-frame fluxes (or 90\% upper-limits), 
$\alpha_{ox}$, a flag indicating the best-fit X-ray spectral model, the photon index, intrinsic 
absorption (or 90\% upper-limit) and the $\chi^2$ values and degrees of freedom for the best-fit 
model.  The best-fit flag indicates which values are listed for each best-fit model.  For sources 
that prefer the APL model (flag = 3), both $\Gamma$ and N$_H$ are from the APL fit.  If the FPL 
model is preferred (flag = 2), the SPL $\Gamma$ and FPL N$_H$ are listed.  For sources that prefer 
the SPL model (flag = 1), $\Gamma$ is from the SPL fit and N$_H$ is the 90\% upper-limit from the 
APL fit.

\section{Results and Discussion}

The \emph{SDSS/XMM-Newton} Quasar Survey contains 792 sources, 685 of which are detected 
in the X-rays and 473 of which have X-ray spectra.  (All have optical spectra.)  The catalog 
covers redshifts z = 0.11 - 5.41 and optical magnitudes range from $i$ = 15.3 to $i$ = 20.7.  
Figures 4 shows the survey sensitivity in the optical ($a$) and X-ray ($b$) bands, and 
the observed-frame 2-10 keV flux distribution ($c$).  The FWHM of the distribution spans 
F$_{2-10keV}$ = (1 - 10) x 10$^{-14}$ ergs cm$^{-2}$ s$^{-1}$ and contains $\sim$ 60\% of the 
detected sources.  

Results of standard $\alpha_{ox}$ analysis (i.e. using the conventional 2500 $\mbox{\AA}$ and 2 keV 
fiducial points) are shown in Figure~\ref{fig:aox}.  We test for correlations with $\alpha_{ox}$ using 
the Kendall correlation test available in ASURV, a survival statistics package \citep{Lavalley92}.  
Figure 5$a$ shows log L$_{2keV}$ vs. log L$_{2500}$ with a dotted line of slope unity for reference.  
The best-fit line, obtained from the EM (estimate and maximize) algorithm within ASURV, is flatter 
than unity with a slope of 0.64 $\pm$  0.03.  This deviation from unity is clear in Figure 5$b$, 
which shows the $\alpha_{ox}-l_{2500\AA}$ correlation, significant at the 11.4$\sigma$ level in the 
\emph{SDSS/XMM} Quasar Survey.  The solid line is the best-fit EM regression line for our data: 
\begin{displaymath} \alpha_{ox} = (3.080 \pm 0.376) + (-0.153 \pm 0.012) log(L_{2500\AA}) \end{displaymath}  
The slope is consistent within errors with the best-fit regression from \citet{Steffen06} (dotted 
line in Fig. 5b).  

Figure 5$c$ plots $\alpha_{ox}$ against redshift.  We find no correlation between $\alpha_{ox}$ and 
redshift, as in R05\nocite{Risaliti05}.  The lack of an $\alpha_{ox}-z$ correlation agrees with some 
previous studies \citep{Vignali03,Strateva05,Risaliti05,Steffen06} but not all \citep{Bechtold03,Shen06,Kelly07}.  
In particular, \citet{Kelly07} apply a more sophisticated statistical analysis by allowing for non-linear 
fits with multiple variables.  As a result, they find that $\alpha_{ox}$ depends on both $l_{2500}$ and 
redshift so that quasars become more X-ray loud at low luminosities and higher redshifts.  We have applied 
only linear regressions in this paper, and will apply more complex statistics in a later publication.

The weighted mean of the X-ray spectral slope for the 473 sources with an X-ray S/N $\geq$ 6 is 
$<\Gamma>$ = 1.91$\pm$0.08, with a standard deviation of 0.40, which is consistent with previous 
results \citep[e.g.,][]{Shen06}.  The typical 1$\sigma$ error on $\Gamma$ is 0.15, resulting in 
an intrinsic dispersion $\sigma_{\Gamma}$ = 0.37.  Figure \ref{fig:HistaoxGamma} shows the $\alpha_{ox}$ 
and $\Gamma$ distributions.

We calculate the percentage of sources with significant intrinsic absorption by testing whether 
sources with S/N $\geq$ 6 prefer the APL model over the SPL model, with F-test probability P$_F$ $>$ 
0.95 and acceptable $\chi^2_\nu$, resulting in 34 (7.2\%) absorbed sources.  However, this method 
is biased against low S/N sources, since the APL model has more parameters than the SPL model.  
Since the FPL and SPL models have the same number of parameters, we count sources as absorbed if they 
prefer either the FPL or APL models with P$_F$ $>$ 0.95 and acceptable $\chi^2_\nu$.  This 
method gives 55 sources (11.6\%) that are intrinsically absorbed, which is comparable to the percentage 
found among Type 1 quasars in previous surveys \citep{Mateos05,Green08}.  The XMM-Newton COSMOS survey 
obtained a higher percentage \citep[20\%,][]{Mainieri07} using a lower confidence threshold (P$_F$ $>$ 0.9) 
when applying the F-test.  When we redo our method using the same confidence threshold, we find 82 absorbed 
sources (17.4\%).  

Nevertheless, the amount of absorption in this sample is only a lower-limit for two reasons.  
First, undetected absorption may still exist, particularly in low S/N sources.  For example, 
in the correlation plots (e.g. Figures 7,9 and 11), three sources can be seen with $\Gamma 
\sim$ 0.5.  These sources do not formally prefer either absorption model, but their X-ray 
spectra are cut off at soft energies, suggesting absorption as the likely cause of flat spectra.  
Second, we cannot test for absorption in sources with S/N $<$ 6, and it is possible that these 
low S/N sources have a higher percentage of absorption.  

For the 55 sources that prefer absorption , we calculate the weighted mean of the intrinsic 
column density: N$_H$ = 1.5$\pm$0.3 x 10$^{21}$ cm$^{-2}$.  The absorbed population consists of 6 
broad absorption line objects (BAL), 7 radio-loud (RL), and 42 radio-quiet (RQ), non-BAL objects.  

Table \ref{table:KS} summarizes the weighted means of $\alpha_{ox}$, $\Gamma$ and N$_H$ for the 
three sub-sets of quasar populations: RQ+non-BAL, RL+non-BAL, and BAL.  (There are four quasars 
that are both RL and BALs.)  Because of the small numbers of absorbed spectra in the sub-populations, 
comparing the respective N$_H$ distributions is not meaningful, but $\alpha_{ox}$ and $\Gamma$, 
Kolmogorov-Smirnov (K-S) tests show that the RQ distribution is significantly different from the 
RL distribution, with probabilities P$_{KS}$ $<$ 0.5\% that the two samples are drawn from the same parent 
distribution (Fig. \ref{fig:HistaoxGamma}).  RL quasars are known to be brighter in the X-rays for a 
given 2500 $\mbox{\AA}$ luminosity \citep[e.g.][]{Zamorani81}.  However, while RL quasars are also known 
to have flatter X-ray slopes than RQ quasars \citep{WE87,Williams92,Reeves97,RT00,Page05,Piconcelli05}, 
the average $\Gamma$ value for RL quasars in the \emph{SDSS/XMM} sample ($<\Gamma>$ = 1.85 $\pm$ 0.04) 
is steeper than that found in previous samples ($<\Gamma>$ $\sim$ 1.5-1.75).  Figure 6$b$ shows that 
RL quasars do follow the same Gaussian distribution as RQ quasars for $\Gamma <$ 1.9, but for $\Gamma >$ 
1.9, the RL distribution falls off rapidly.  Only 7 of 49 RL quasars (14\%) have $\Gamma >$ 2, compared to 
48\% of RQ quasars.   

The $\Gamma$ distribution is not significantly different for BAL vs. non-BAL quasars (P$_{KS}$ = 75\%) 
once the X-ray spectra are corrected for intrinsic absorption.  However, the $\alpha_{ox}$ distribution 
of BAL quasars, which includes 37 sources with S/N $<$ 6 that cannot be corrected for absorption, 
remains significantly different (P$_{KS}$ = 0.2\%) from non-BAL quasars.  
Previous studies have noted the difference in X-ray brightness for BAL quasars \citep[e.g.][]{GM96}.

\subsection{Correlations with the X-ray Spectral Slope}

The large number of X-ray spectra in the \emph{SDSS/XMM} Quasar Survey allows us to test for 
correlations with the X-ray photon index ($\Gamma$).  Since we use only sources detected with 
a high enough S/N to fit a power-law, we do not use survival statistics and instead use the 
Kendall's rank correlation coefficient to test for correlations.  We use only those sources 
that do not prefer either absorption model, and we also require that the SPL model result in 
a reasonably good fit ($\chi^2_\nu < 1.2$).  In addition, we restrict the selection to those 
sources with S/N $>$ 10 in order to reduce the effect of undetected absorption in lower S/N 
spectra.  

When correlations are significant, we plot a Weighted Least Squares (WLS) regression line to 
take into account the measurement errors in $\Gamma$, which are much larger than the errors in 
the independent variable.  RQ quasars show two correlations of $\Gamma$ with luminosity and 
optical color with probabilities less than 0.5\% that they are due to chance: 
(1) \emph{$\Gamma$ vs. log L$_{2keV}$} (P$_K$ = 2.4e-5) and 
(2) \emph{$\Gamma$ vs. $\Delta$($g - i$)} (P$_K$ = 7.5e-4), 
where P$_K$ is the Kendall rank two-sided significance level for each correlation.  
RQ quasars also show a marginal correlation between $\Gamma$ and $\alpha_{ox}$ (P$_K$ = 1.6\%).  
The $\Gamma$ - L$_{2keV}$, $\Gamma$ - $\Delta$($g - i$) and $\Gamma$ - $\alpha_{ox}$ relations 
are not significant (P$_K >$ 10\%) for RL quasars (green stars in Figures 7, 9 and 11).  
We now discuss each of these correlations in turn.  

\subsubsection{X-ray Slope vs. X-ray Luminosity}

An anti-correlation exists between the X-ray slope ($\Gamma$) and the 2 keV luminosity 
(log L$_{2keV}$), hereafter $l_{2keV}$, such that X-ray slopes harden as X-ray luminosity 
increases (Fig. 7$a$).  The WLS regression line (for RQ, non-BAL quasars) is: 
\begin{equation}\label{GLx} \Gamma = (6.39 \pm 1.3) + (-0.16 \pm 0.05) l_{2keV} \end{equation}
To investigate the possibility of a pivot point in the X-ray spectrum, we next examined 7 
additional correlations between the X-ray slope ($\Gamma$) and the monochromatic X-ray luminosities at 
0.7, 1, 1.5, 4, 7, 10 and 20 keV.  The monochromatic fluxes are obtained for sources preferring 
the SPL model by normalizing the fit at each energy in turn.  The results of these correlations 
are summarized in Figures 7$a-d$ and in Table \ref{table:GLx}.  The strength of the correlation 
increases with energy as the slope steepens, so the strongest, steepest correlation is between 
$\Gamma$ and $l_{20keV}$ (Fig. 7$d$).  Correlation strength decreases at lower energies, bottoming 
out at 1 keV, where the slope is consistent with zero.  At 0.7 keV, the slope flips to a positive 
correlation and the strength of the correlation increases again.  The flip in the correlation slope 
indicates a pivot in the X-ray spectrum near 1 keV (Fig. 8$a$).  

For the correlations above, we have used $\Gamma$ to extrapolate the rest-frame monochromatic 
luminosities for sources with redshifts out of range of the observed spectrum.  To check that this 
extrapolation does not affect the correlations, we perform the fits again, this time excluding any 
sources where the rest-frame luminosity is not in the observed range.  The results are shown in Table 
5, and plotted in Fig. 8$b$.  The errors are larger, due to the smaller sample sizes and the narrower 
range of luminosities observed, but the spectrum still pivots between 1 and 1.5 keV.  At the lowest 
energies, the slopes become much steeper, possibly due to the influence of a soft excess component 
on the power-law fit.  

The $\Gamma$ determined from the SPL model is used to determine the monochromatic flux, so it 
is important to check that the model assumptions do not induce the observed anti-correlation.  
To do this, we re-fit the sources, this time with a power-law fixed to the sample mean ($\Gamma$ 
= 1.91) and intrinsic absorption fixed to zero.  We then obtain the monochromatic flux at 2 keV 
and compare the F$_{2keV}$ obtained via a fixed power-law to the F$_{2keV}$ obtained via the 
best-fit power-law.  We find that changing the model assumptions changes the log flux values by 
1.4\%, which is not enough to explain the observed correlation at 2 keV.  

Even with the selection restricted to sources with S/N $>$ 10, there is still the possibility of 
undetected absorption.  For example, Figures 7 show two sources with $\Gamma$ $\sim$ 0.5 and 
another four radio-quiet sources with $\Gamma$ $\sim$ 1-1.5.  The X-ray spectra of these sources 
show curvature in the soft X-rays that, while not significant enough for the sources to prefer an 
absorption model, nevertheless suggests  that intrinsic absorption is the likely cause of flat 
X-ray slopes.  Therefore, we test for correlations again, this time using sources with S/N $>$ 20 
in order to minimize the chance of undetected absorption.  The significance of the 
correlation increases, with the probability of a chance correlation falling below P$_K$ = 1e-7.  
However, with many low-luminosity sources eliminated, the correlation is biased to a steeper 
slope.  Having shown that the correlations do not rely on undetected absorption, we continue 
to use the correlations for sources with S/N $>$ 10 for further analysis.  

The slope of the $\Gamma-l_{2keV}$ anti-correlation is equivalent within errors to the slope found in 
\citet{Green08}.  They also find a correlation between $\Gamma$ and the 2 keV luminosity based on 
156 RQ and RL quasars fit with a power-law plus intrinsic absorption, plus 979 quasars with lower 
S/N that were fit with a single power-law and zero intrinsic absorption.  A similar anti-correlation 
between $\Gamma$ and 2-10 keV luminosity was found in \citet{Page05} for 16 RL quasars.  While the 
\citet{Page05} correlation was not found to be significant for RQ quasars, there were only 7 RQ 
quasars in their sample.  However, a previous study by \citet{Dai04} found the opposite correlation 
between $\Gamma$ and $l_{2-10keV}$ at 98.6\% significance for a sample of 10 quasars observed with 
\emph{Chandra} and \emph{XMM}.  The $\Gamma$ and extrapolated $l_{2keV}$ values for the \citet{Dai04} 
quasars are plotted in Fig. 7$a$ as open, black squares, where they are consistent with the trend 
observed in \emph{SDSS-XMM} quasars.\footnote{More recently, \citet{Saez08} find the same correlation 
to be significant to $>$ 99.5\% for 173 bright, radio-quiet quasars in the \emph{Chandra} Deep Fields.  
However, the trends found in \citet{Saez08} are dominated by Type 2 quasars, particularly at low X-ray 
luminosities.  As we have excluded Type 2 quasars, the two samples are not in conflict.}  The 
SDSS/XMM-Newton Quasar Survey increases the sample size of previous studies by factors of 3 - 30, and 
covers $\sim3$ decades of X-ray luminosity.  

To explain the observed correlations between $\Gamma$ and X-ray luminosity, we must answer two questions:

1.  Why does the X-ray slope change with luminosity?

2.  Why does the slope change such that the pivot point is near 1 keV?

We address two possible answers to the first question.  First, an increased hard component at higher 
X-ray luminosities may explain the trends observed here.  The hard component may be due to nonthermal 
emission associated with a jet \citep[e.g.][]{Zamorani81,WE87} or due to a reflection component 
\citep{Krolik99}.  RL quasars do not show a significant correlation between $\Gamma$ and 
L$_X$ except for marginal correlations at high energies (7 and 10 keV), but this may be because a jet 
already dominates the X-ray emission.  The spectrum of the hard component may flatten the X-ray spectrum 
by covering up the steeper power-law due to inverse Compton scattering, while simultaneously increasing 
the 2-10 keV luminosity.  

Alternatively, quasars with low X-ray luminosities may have steeper slopes due to a component linked to 
high accretion rates.  As discussed in $\S$1.1, previous studies have found a strong correlation between 
$\Gamma$ and the Eddington ratio (L$_{bol}$/L$_{Edd}$), where steeper sources are associated with higher 
accretion rates.  Therefore, sources with high accretion rates (and steep X-ray spectra) must be associated 
with low X-ray luminosity in order to produce the $\Gamma$-L$_X$ correlations.  However, it is not clear 
if high accretion rates and steep X-ray spectra tend to coincide with low X-ray luminosity.
Studies of X-ray binary (XRB) accretion states have shown that high accretion states tend to be associated 
with steep X-ray spectra and \emph{high} X-ray flux compared to the low accretion state \citep{RMcC06}.  
However, the distinction is not clear in every source - even "high" accretion states can be 
associated with both low and high X-ray flux.  Narrow-line Seyfert 1 (NLS1) galaxies are an extreme example 
of high accretion rate objects \citep{Boroson02}, but while they typically have steep X-ray slopes, their 
X-ray flux can range from X-ray bright to weak \citep[e.g., $\alpha_{ox}$ = 0.7-2.2][]{Leighly99}.
The $\Gamma$-L$_{bol}$/L$_{Edd}$ relation for the SDSS/XMM-Newton Quasar Survey will be discussed further 
in (Risaliti et al. 2009, in prep.)\nocite{Risaliti09}.

Inverse Compton scattering of UV photons from an accretion disk in a hot corona could explain why the X-ray 
spectrum pivots at low X-ray energies.  In the corona model, the X-ray spectrum will change shape if the 
temperature (T$_e$) and/or optical depth ($\tau$) of the corona vary \citep{RL79}, since the output spectrum 
flattens as the $y$ parameter increases.  The $y$ parameter is the average fractional energy gained by a photon; 
for a thermal, non-relativistic electron distribution, 
\begin{equation} y = \frac{4kT_e}{m_ec^2} max(\tau,\tau^2) \end{equation}
For example, if the disk emission brightens, increasing the soft photon supply, the corona will 
increase radiative cooling to maintain temperature balance, producing a steeper X-ray spectrum.  For 
a T $\sim$ 10$^8$-10$^9$ K corona, opacity is not necessarily dependent on luminosity, so opacity 
variations can result in a pivot in the 2-10 keV band without a large accompanying change in luminosity 
\citep{Haardt97}.  

Because of the relation between the physical parameters and the output spectrum, it is possible to 
calculate T$_e\tau$, though the degeneracy is not breakable without knowing the cut-off of the 
high-energy spectrum.  From \citet[p. 227-8, ][]{Krolik99}, 
\[\ T_e\tau \sim a ~ (l_h/l_s)^{1/4}  \sim a ~ \frac{-1.6}{\Gamma-1}, \]
where $a$ is a coefficient dependent on geometry ($a$ = 0.06 for slabs) and $l_h/l_s$ is the compactness 
ratio, equivalent to the heating rate over the soft photon seed supply.  
Therefore, as $\Gamma$ changes from 2.3 at low luminosities to 1.8 at high luminosities, T$_e\tau$ 
approximately doubles from 4.16 x 10$^8$ to 7.13 x 10$^8$ K.  

Several studies have found evidence of pivot points in individual objects.  The previous paragraph describes 
the model of Mrk766, a NLS1, which was found to pivot near $\sim$10 keV \citep{Haardt97}.  In Cygnus X-1, 
a black hole X-ray binary, \citet{Zdziarski02} find a negative correlation between X-ray flux and $\Gamma$ 
for the 0.15-12 keV band; no correlation for the 20-100 keV band; and a positive correlation for flux greater 
than 100 keV, implying a pivot point near $\sim$50 keV.  \citet{Zdziarski02} model this spectrum as a variable 
supply of soft seed photons irradiating a thermal plasma, where optical depth is constant.  An increase in 
seed flux results in a decrease in the corona temperature to satisfy energy balance, resulting in a steeper 
spectrum with a pivot somewhere below 100 keV.  Two model-independent analyses of the spectral variability in 
Seyfert 1 galaxies also show evidence of pivot points.  NGC 4051 was found to have a pivot near 50 keV using 
correlations between the 2-5 keV and 7-15 keV fluxes \citep{Taylor03}, and NGC 5548, studied with an 8-day 
BeppoSAX observation \citep{Nicastro00}, can be entirely explained by a pivot in the medium energy band 
($\sim$6 keV).  While these examples show that pivot points may exist in X-ray spectra, it is clear that 
further exploration must be done to provide a clear picture of the underlying physics.

\subsubsection{X-ray slope vs. optical color and $\alpha_{ox}$ (for RQ quasars only)}

X-ray slope ($\Gamma$) and optical color [$\Delta$($g - i$)] are correlated at 3.4$\sigma$, 
such that quasars with redder colors are more likely to have flat X-ray slopes (Fig. 9$a$).  
This is likely due to undetected absorption, which is discussed further below.  We exclude 
sources with redshifts greater than 2.3 from the $\Gamma$ - $\Delta$($g - i$) correlation as 
Ly$\alpha$ absorption artificially reddens high-redshift quasars.  The linear bisector is: 
\begin{equation}\label{Ggi} \Gamma = (2.04 \pm 0.02) + (-0.45 \pm 0.10)  \Delta(g - i) \end{equation}  
The $\Gamma$ - $\Delta$($g - i$) correlation is consistent with previous findings of a correlation 
between $\Gamma$ and the optical/UV spectral slope ($\alpha_{uv}$) \citep{Kelly07} because color 
correlates tightly with the optical slope \citep{Richards03} ($\Delta$($g - i$) $\propto$ 0.5
$\alpha_{uv}$).  The slope of the correlation in this paper is in the same direction as, but 
steeper than in \citet{Kelly07}, who find a slope = -0.25 $\pm$0.07.

X-ray slope ($\Gamma$) anticorrelates with $\alpha_{ox}$ at marginal (2.4$\sigma$) significance, 
such that X-ray faint quasars are more likely to have flat X-ray slopes (Fig. 9$b$).  
The WLS regression line is: 
\begin{equation}\label{Gaox} \Gamma = (2.8 \pm 0.2) + (0.6 \pm 0.1)  \alpha_{ox} \end{equation}
This anti-correlation between $\Gamma$ and $\alpha_{ox}$ contrasts with the slightly positive 
correlation found by \citet{Green08} (overplotted as the almost horizontal dashed line in Fig. 8$b$).  
The \citet{Green08} correlation may be affected by the inclusion of RL quasars, which lie at the 
X-ray bright end and typically have flatter X-ray slopes than RQ quasars.

Completing the triangle of relations, a correlation is found between $\alpha_{ox}$ and $\Delta$($g - i$) 
at 4.7$\sigma$ significance (Fig. 9$c$).  Again, sources with redshifts greater than 2.3 are excluded in 
order to avoid contamination by the Ly$\alpha$ forest.  Since the $\alpha_{ox}$-$\Delta$($g - i$) relation 
includes censored data, we use the EM method to give the best-fit line:  
\begin{equation}\label{aoxgi} \alpha_{ox} = (-1.613 \pm 0.009) + (-0.16 \pm 0.04)\Delta(g - i) \end{equation}  
The EM algorithm assumes that all of  the error lies in the dependent variable, so the algorithm reduces 
to an OLS Y vs. X fit for uncensored data.  

In an attempt to disentangle the relationships between $\Gamma$, $\Delta$($g - i$), and $\alpha_{ox}$, 
we perform the Kendall partial correlation test for sources with S/N $>$ 6 and z $<$ 2.3, but each 
correlation is significant at the $\sim$3$\sigma$ level even when the third variable is taken into 
account.  The dispersions in all three relations are large, so the sources are binned along the x-axis 
to show the relationships more clearly.  The binning shows that the relations are dominated by outliers: 
optically red and X-ray weak/flat sources.  

Detected absorption cannot explain the observed correlations with $\Gamma$, since sources are only 
included if they do not prefer a model with intrinsic X-ray absorption.  $\Gamma$ is not 
correlated with N$_H$ (Fig. 10$a$) for sources that prefer the APL model, indicating that once the X-ray 
slope is corrected for absorption, no intrinsic correlation between X-ray slope and absorption remains.  
For the same sources, $\alpha_{ox}$ and N$_H$ (Fig. 10$b$) correlate at the 3.3$\sigma$ level, which is 
surprising since the 2 keV flux should also be corrected for absorption.  Since the detections trace the 
upper-limits, this correlation is likely not intrinsic, but is due to the survey flux limit.  
The optical color also depends on absorption (Fig. 10$c$) at the 2.6$\sigma$ level, as is expected, since 
X-ray absorbed sources are more likely to have redder colors due to dust-reddening.  

Undetected absorption is a possible explanation for all three relations, causing red optical colors 
and X-ray weakness while flattening the X-ray spectral slope in low S/N spectra.  A simple calculation 
shows that the amounts of absorption obtained from equations (\ref{Ggi}) and (\ref{Gaox}) are consistent 
with observed gas-to-dust ratios observed for quasars.  First, we assume that both relations are driven 
entirely by the effects of absorption (i.e. ignoring any possible effects due to intrinsic properties, 
such as black hole mass and accretion rate).  We also assume that an unabsorbed quasar has a typical 
X-ray slope, ($\Gamma$ = 1.9), zero absorption (N$_H$ = 0 and E$_{B-V}$ = 0), and typical blue optical 
colors ($\Delta$($g - i$) = 0).  If the X-ray slope changes by a given amount, we can calculate the change 
in color by equation (\ref{Ggi}), which corresponds to a dust-reddening \citep{Richards03}, which in turn 
leads to a change in optical luminosity.  Since the X-ray slope also gives $\alpha_{ox}$ by equation 
(\ref{Gaox}), we can calculate the change in X-ray luminosity at 2 keV, which in turn gives the intrinsic 
X-ray absorption.  So for an absorbed quasar with $\Gamma$ = 1.4, the dust-reddening is E(B-V) = 0.096 
from equation (\ref{Ggi}), the gas column is N$_H$ $\sim$ 6 x 10$^{22}$ cm$^{-2}$ from equation (\ref{Gaox}), 
giving a gas-to-dust ratio $\sim$ 100 times the Galactic value.  Quasars typically have gas-to-dust ratios 
in the range of 10-100 times the Galactic value \citep{Maccacaro81, Maiolino01, Wilkes02}, so relations 
(\ref{Ggi}) and (\ref{Gaox}) are consistent with intrinsic absorption that remains undetected in the X-ray 
spectra.  

Intrinsic absorption is less likely to remain undetected in spectra with high S/N.  As a second test, we restrict 
the correlation tests to those sources with X-ray S/N $>$ 20.  As a result, all three correlations disappear, 
which again supports undetected absorption as the cause of correlations that include low S/N sources.  

\subsubsection{X-ray slope vs. redshift and optical luminosity}

Neither redshift nor 2500 $\mbox{\AA}$ luminosity correlate significantly with $\Gamma$ ($P_S >$ 10\%, Figures 
11), confirming previous studies with smaller samples \citep{Page04,Risaliti05,Shemmer05,Vignali05,Kelly07}.  
Note however \citet{Bechtold03}, who find that X-ray slopes are flatter at lower redshifts for a sample of 
17 radio-quiet, high-redshift (3.7 $<$ z $<$ 6.3) quasars observed with \emph{Chandra}.  
The \emph{SDSS/XMM} sample (473 sources with X-ray spectra) is at least an order of magnitude larger than 
these previous samples, though it does not have homogenous spectral coverage for redshifts above 2.5.  

\section{Conclusions and Future Work}

We have cross-correlated the DR5 SDSS Quasar Catalog with the \emph{XMM-Newton} archive, creating a sample 
of 792 quasars with a detection rate of 87\%.  Almost 500 quasars have X-ray spectra, the largest sample 
available for analysis of optical/X-ray spectral correlations.  We find that the X-ray photon index $\Gamma$ 
correlates significantly with $l_{X}$, where X = 2, 4, 7, 10 and 20 keV.  
Optical color and $\alpha_{ox}$ also correlate with $\Gamma$, but these correlations are likely due to the 
effect undetected intrinsic absorption rather than intrinsic physical changes.  
The X-ray slope does not correlate significantly with redshift or optical luminosity.  With a sample size 
at least an order of magnitude larger than previous studies, we confirm a highly significant correlation 
between $\alpha_{ox}$ and the monochromatic luminosity at 2500 $\mbox{\AA}$, and we also confirm that 
$\alpha_{ox}$ does not correlate significantly with redshift or optical color.  

Future studies of the sample will include: $\alpha_{ox}$, $\Gamma$ vs optical properties, and studies of 
sub-populations such as BALs and Type 2 quasars.  Variability studies will also be pursued for 265 objects with 
multiple \emph{XMM} observations.  More complex spectral analysis on high X-ray S/N sources will include 
a thermal component for the soft excess, warm absorbers, and emission line detection.

\acknowledgements
The authors thank the anonymous referee for insightful comments that improved this paper.  
The authors also thank Gordon Richards for his excellent assistance in navigating the SDSS database.  
This paper is based on observations obtained with XMM-Newton, an ESA science mission with 
instruments and contributions directly funded by ESA Member States and NASA, and the Sloan 
Digital Sky Survey (SDSS).  Funding for the SDSS and SDSS-II has been provided by the 
Alfred P.  Sloan Foundation, the Participating Institutions, the National Science Foundation, 
the U.S.  Department of Energy, the National Aeronautics and Space Administration, the 
Japanese Monbukagakusho, the Max Planck Society, and the Higher Education Funding Council 
for England.  This research also made use of the NASA/ IPAC Infrared Science Archive, 
which is operated by the Jet Propulsion Laboratory, California Institute of Technology, 
under contract with the National Aeronautics and Space Administration.  This work has 
been partially funded by NASA Grants NASA NNX07AI22G and NASA GO6-7102X. 

\clearpage


\section{Erratum: The Fifth Data Release Sloan Digital Sky Survey/XMM-NEWTON Quasar Survey (2009, ApJS, 183, 17); published Nov. 3, 2009}

We have discovered an error in column 9 of Table 2 in the original paper.  This 
column reports the fit flags, indicating which model a source prefers: a simple 
power-law (SPL, flag = 1), a fixed power-law plus intrinsic absorption (FPL, flag 
= 2), or an absorbed power-law (APL, flag = 3).  The table of the original paper 
mistakenly reports all flag=3 sources as having flag=2, and all flag=2 sources as 
having flag=1, so that only 32 sources prefer an absorption model.  

We have updated Table 2 to print out the correct fit flags for each source, resulting 
in 55 sources that prefer an absorption model.  Since the fit flags determine which numbers 
are reported for the remaining columns of Table 2, these numbers are updated as well.  
The abstract and text of the original paper report the correct number of absorbed sources, 
so the conclusions are unaffected. 

In addition, a minor rounding error was found in the SDSS names of some objects, and so 
we replace both Table 1 and 2 with corrected versions.  Corrected machine-readable tables 
are available online from ApJS and will soon be available on the High Energy Astrophysics 
Science Archive Resource Center (HEASARC, http://heasarc.gsfc.nasa.gov).

\begin{deluxetable}{lcccccccccc}
\tabletypesize{\footnotesize}
\tablecolumns{11} 
\tablewidth{0pt} 
\tablecaption{SDSS quasars with archival XMM-Newton observations \label{table:Xray obs}}
\tablehead{\colhead{SDSS name}                               &
           \colhead{XMM ID}                                 &
           \colhead{z}                                      &
           \colhead{N$_{H,gal}$\tablenotemark{a}}           &
           \colhead{(S/N)$_X$\tablenotemark{b}}             &
           \colhead{Exp.}		            	    &	
	   \colhead{$\Theta$\tablenotemark{d}}		    &
           \colhead{Net}                                    &
           \colhead{Background}                             &
	   \colhead{RL}                                     &
           \colhead{BAL}                                    \\
	   \colhead{}                                       &
           \colhead{}                                       &
           \colhead{}                                       &
           \colhead{}                                       &
           \colhead{}                                       &
           \colhead{time\tablenotemark{c}}                  &
           \colhead{}                                       &
           \colhead{counts\tablenotemark{e}}                &
           \colhead{counts\tablenotemark{f}}                &
           \colhead{flag\tablenotemark{g}}                  &
           \colhead{flag\tablenotemark{h}}                  }
\startdata
SDSS J125930.97+282705.5  & 0204040101 & 1.094 & 0.92 & 18.0 & 221.0 & 13.4 &  469 & 103.  & 0 & 0\\
SDSS J130120.13+282137.2  & 0204040101 & 1.369 & 0.94 & 45.8 & 221.0 & 13.2 & 2780 & 449.  & 1 & 0\\
SDSS J114856.56+525425.2  & 0204260101 & 1.633 & 1.37 & 23.0 &   9.4 & 5.8  &  620 & 52.1  & 1 & 0\\
SDSS J215419.70-091753.6  & 0204310101 & 1.212 & 3.71 & 12.5 &  81.3 & 10.9 &  218 & 43.5  & 0 & 0\\
SDSS J164221.22+390333.4  & 0204340101 & 1.713 & 1.22 & 11.3 &  45.7 & 11.9 &  182 & 39.1  & 0 & 0\\
SDSS J021000.22-100354.2  & 0204340201 & 1.960 & 2.20 & 13.2 &  31.6 & 10.9 &  243 & 47.0  & 1 & 0\\
SDSS J021100.83-095138.4  & 0204340201 & 0.767 & 2.17 & 14.4 &  31.6 & 11.8 &  268 & 40.6  & 0 & 0\\
SDSS J123508.19+393020.0  & 0204400101 & 0.968 & 1.49 &  9.9 &  65.5 & 4.8  &  153 & 42.8  & 0 & 0\\
SDSS J123527.36+392824.0  & 0204400101 & 2.158 & 1.49 & 20.1 &  89.8 & 6.0  &  553 & 104.  & 0 & 0\\
SDSS J133812.97+391527.1  & 0204651101 & 0.439 & 0.86 & 18.7 &  23.1 & 8.0  &  416 & 38.4  & 0 & 0\\
\enddata

\tablenotetext{a}{Galactic hydrogen column density (10$^{20}$ cm$^{-2}$) in the direction of the source.}

\tablenotetext{b}{X-ray signal-to-noise}

\tablenotetext{c}{Exposure time (kiloseconds)}

\tablenotetext{d}{Off-axis angle (arcminutes)}

\tablenotetext{e}{Net source counts}

\tablenotetext{f}{Background counts (counts in background region scaled to source extraction area)}

\tablenotetext{g}{Radio-loud flag is 0 for RQ quasars, 1 for RL quasars, and 2 if the radio upper-limit is 
too high to determine whether source is RL.  }

\tablenotetext{h}{Broad-absorption line flag is 0 for non-BAL quasars, 1 for BAL quasars}

\tablecomments{Table~\ref{table:Xray obs} is presented in its entirety in the electronic edition of the Astrophysical Journal. 
A portion is shown here for guidance regarding its form and content.  The upper limits are at the 1.6$\sigma$ level.}

\end{deluxetable}

\begin{deluxetable}{lccccccccccc}
\tabletypesize{\footnotesize}
\tablecolumns{12} 
\tablewidth{0pt} 
\tablecaption{X-ray Spectral Data of SDSS Quasars\label{table:specdata}}
\tablehead{\colhead{SDSSname}                    	&
           \colhead{f$_{0.5-2}$\tablenotemark{a}}     	&
           \colhead{f$_{2-10}$\tablenotemark{a}}      	&
           \colhead{f$_{0.5-2}$\tablenotemark{b}}     	&
           \colhead{f$_{2-10}$\tablenotemark{b}}      	&
           \colhead{L$_{0.5-2}$\tablenotemark{c}}     	&
           \colhead{L$_{2-10}$\tablenotemark{c}}	&
           \colhead{$\alpha_{ox}$}			&
           \colhead{Fit}			        &      
           \colhead{$\Gamma$\tablenotemark{e}}          &
           \colhead{N$_H$\tablenotemark{f}}             &
           \colhead{$\chi^2 / \nu$\tablenotemark{g}}    \\
	   \colhead{}					&
	   \colhead{(obs.)}				&
	   \colhead{(obs.)}				&
	   \colhead{(rest)}				&
	   \colhead{(rest)}				&
	   \colhead{(rest)}				&
	   \colhead{(rest)}				&
	   \colhead{}					&
	   \colhead{flag\tablenotemark{d}}		&
	   \colhead{}					&
	   \colhead{}					&
	   \colhead{}					}

\startdata
SDSS J125930.97+282705.5  & 0.94 & 5.20 & 0.41 & 2.46 & 0.27 & 1.61 & -1.77 & 1 & 0.99$_{-0.14}^{+0.13}$ & $<$3.10 & 38.9/28 \\
SDSS J130120.13+282137.2  & 12.4 & 20.5 & 9.01 & 17.0 & 10.3 & 19.3 & -1.32 & 1 & 1.78$_{-0.07}^{+0.07}$ & $<$0.05 & 90.0/120 \\
SDSS J114856.56+525425.2  & 26.3 & 59.9 & 14.0 & 39.5 & 24.5 & 69.4 & -1.60 & 1 & 1.58$_{-0.12}^{+0.12}$ & $<$0.73 & 20.3/26 \\
SDSS J215419.70-091753.6  & 1.51 & 2.38 & 0.93 & 2.10 & 0.79 & 1.77 & -2.12 & 1 & 1.87$_{-0.27}^{+0.29}$ & $<$0.68 & 5.0/12 \\
SDSS J164221.22+390333.4  & 3.09 & 6.23 & 1.76 & 4.39 & 3.48 & 8.68 & -1.52 & 1 & 1.66$_{-0.24}^{+0.25}$ & $<$0.44 & 4.5/12 \\
SDSS J021000.22-100354.2  & 5.63 & 12.7 & 2.35 & 8.04 & 6.49 & 22.1 & -1.37 & 1 & 1.60$_{-0.19}^{+0.19}$ & $<$1.38 & 7.9/14 \\
SDSS J021100.83-095138.4  & 5.72 & 5.04 & 5.72 & 5.73 & 1.56 & 1.56 & -1.41 & 1 & 2.23$_{-0.22}^{+0.24}$ & $<$0.19 & 5.6/16 \\
SDSS J123508.19+393020.0  & 1.68 & 1.45 & 1.74 & 1.69 & 0.84 & 0.82 & -1.52 & 1 & 2.23$_{-0.33}^{+0.38}$ & $<$0.65 & 5.1/13 \\
SDSS J123527.36+392824.0  & 2.57 & 3.40 & 1.57 & 3.13 & 5.49 & 10.9 & -1.51 & 1 & 1.94$_{-0.16}^{+0.17}$ & $<$0.54 & 20.7/25 \\
SDSS J133812.97+391527.1  & 9.13 & 38.0 & 6.69 & 28.2 & 0.47 & 1.98 & -1.50 & 1 & 1.17$_{-0.16}^{+0.17}$ & $<$0.08 & 14.1/22 \\
\enddata

\tablenotetext{a}{Observed-frame X-ray flux in the observed band  is given in units of 10$^{-14}$ ergs cm$^{-2}$ s$^{-1}$.}

\tablenotetext{b}{Rest-frame X-ray flux in the soft and hard bands is given in units of 10$^{-14}$ ergs cm$^{-2}$ s$^{-1}$.}

\tablenotetext{c}{Rest-frame X-ray luminosity in the soft and hard bands is given in units of 10$^{44}$ ergs s$^{-1}$.}

\tablenotetext{d}{A flag indicating the X-ray fit.  An undetected source is flagged as -1 and upper-limit 
flux values are listed.  A detected source with S/N $<$ 6 is flagged as 0 and detected flux values are 
listed.  For sources with S/N $>$ 6, three models can be applied.  If a single power-law model (SPL) is 
preferred, the flag = 1 and the SPL power-law parameters are listed, as well as the 90\% upper-limit on 
intrinsic absorption from the intrinsically absorbed power-law (APL) model.  If a fixed power-law (FPL) 
model with variable N$_H$ is preferred, the flag = 2.  The best-fit slope from the SPL is listed, as well 
as the best-fit N$_H$ from the FPL model.  If the APL model is preferred, the flag = 3, and the APL 
power-law and absorption parameters are listed.}  

\tablenotetext{e}{Photon-index for the best-fit model.  (If the APL model is preferred, then the photon 
index from that model is quoted; otherwise, the photon index is from the single power-law model.)}

\tablenotetext{f}{Intrinsic absorption or 90\% upper-limit in units of 10$^{22}$ cm$^{-2}$.}

\tablenotetext{g}{The $\chi^2$ value and degrees of freedom for the the best-fit model.}




\tablecomments{Table~\ref{table:specdata} is presented in its entirety in the electronic edition of the Astrophysical Journal. 
A portion is shown here for guidance regarding its form and content.Luminosities are computed using H$_0$ = 
70 km s$^{-1}$ Mpc$^{-1}$, $\Omega_M$ = 0.3, and $\Omega_\Lambda$ = 0.7. The upper limits are at the 1.6$\sigma$ level.}
\end{deluxetable}

\begin{deluxetable}{llllllll}
\tabletypesize{\footnotesize}
\tablecolumns{8} 
\tablewidth{0pt} 
\tablecaption{Weighted Averages of X-ray Spectral Quantities\label{table:KS}}
\tablehead{\colhead{}              		 		&
	   \colhead{N$_{det}$\tablenotemark{a}}  		&
	   \colhead{$\alpha_{ox}$} 		 		&
	   \colhead{N$_{spec}$\tablenotemark{b}} 		&
           \colhead{$\Gamma$}      				&
	   \colhead{$\sigma_\Gamma$\tablenotemark{c}}           &
	   \colhead{$\sigma_{\Gamma,intr}$\tablenotemark{d}}    &	
           \colhead{N$_H$\tablenotemark{e}}      		}

\startdata

All		&  685  &  -1.60  &  473  &  1.90$\pm$0.02  & 0.40 & 0.37 &  0.15$\pm$0.03 \\
RQ+non-BAL      &  589  &  -1.61  &  411  &  1.91$\pm$0.08  & 0.41 & 0.38 &  0.14$\pm$0.04 \\
RL+non-BAL	&  62   &  -1.46  &  47   &  1.85$\pm$0.03  & 0.30 & 0.26 &  0.15$\pm$0.03 \\
BAL		&  34   &  -1.78  &  15   &  1.96$\pm$0.05  & 0.37 & 0.34 &  2.3$\pm$0.6   \\

\enddata

\tablenotetext{a}{Number of detected quasars in each sample (S/N $\geq$ 2).}

\tablenotetext{b}{Number of quasars in each sample with X-ray spectra (S/N $\geq$ 6).}

\tablenotetext{c}{Observed dispersion of X-ray slope.}

\tablenotetext{d}{Intrinsic dispersion of X-ray slope}

\tablenotetext{e}{Intrinsic absorption in units of 10$^{22}$ cm$^{-2}$.}
\end{deluxetable}

\begin{deluxetable}{llllll}
\tabletypesize{\footnotesize}
\tablecolumns{6} 
\tablewidth{0pt} 
\tablecaption{$\Gamma$ - L$_X$ Correlations\label{table:GLx}}
\tablehead{\colhead{E (keV)\tablenotemark{a}}              	&
	   \colhead{P$_K$\tablenotemark{b}} 		 	&
	   \colhead{Z-level\tablenotemark{c}} 		 	&
	   \colhead{slope}		 			&
           \colhead{intercept}      				&
	   \colhead{dispersion}           			}

\startdata
0.7	&  0.0062  &  2.73  &   0.095$\pm$0.058  & -0.500$\pm$1.564  &  0.36\\
1.0	&  0.1683  &  0.16  &  -0.015$\pm$0.054  &  2.462$\pm$1.449  &  0.37\\
1.5	&  0.0033  &  1.77  &  -0.106$\pm$0.051  &  4.903$\pm$1.363  &  0.37\\
2.0	&  0.0000  &  2.93  &  -0.162$\pm$0.048  &  6.386$\pm$1.286  &  0.37\\
4.0	&  0.0000  &  5.61  &  -0.265$\pm$0.041  &  9.042$\pm$1.082  &  0.35\\
7.0	&  0.0000  &  7.50  &  -0.316$\pm$0.035  &  10.28$\pm$0.917  &  0.33\\
10.0	&  0.0000  &  8.60  &  -0.335$\pm$0.032  &  10.73$\pm$0.818  &  0.31\\
20.0    &  0.0000  &  9.61  &  -0.365$\pm$0.027  &  11.38$\pm$0.690  &  0.29\\
\enddata

\tablenotetext{a}{Energy at which monochromatic luminosity is taken.}

\tablenotetext{b}{Kendall's probability that correlation is due to chance.}

\tablenotetext{c}{Significance level of Kendall rank correlation coefficient in units of 1$\sigma$.}
\end{deluxetable}

\begin{deluxetable}{llllll}
\tabletypesize{\footnotesize}
\tablecolumns{6} 
\tablewidth{0pt} 
\tablecaption{$\Gamma$ - L$_X$ Correlations with redshift cut-offs}
\tablehead{\colhead{E (keV)\tablenotemark{a}}              	&
	   \colhead{P$_K$\tablenotemark{b}} 		 	&
	   \colhead{sig\tablenotemark{c}} 		 	&
	   \colhead{slope}		 			&
           \colhead{intercept}      				&
	   \colhead{dispersion}           			}
\startdata
0.7	&  0.0112  &  2.54$\sigma$  &   0.583$\pm$0.285  & -13.29$\pm$7.587  &  0.46\\
1.0	&  0.0386  &  1.43$\sigma$  &   0.196$\pm$0.129  &  3.077$\pm$3.447  &  0.44\\
1.5	&  0.0024  &  2.07$\sigma$  &  -0.120$\pm$0.058  &  5.270$\pm$1.363  &  0.38\\
2.0	&  0.0000  &  3.03$\sigma$  &  -0.162$\pm$0.053  &  6.800$\pm$1.417  &  0.38\\
4.0	&  0.0000  &  5.64$\sigma$  &  -0.178$\pm$0.044  &  9.596$\pm$1.176  &  0.35\\
7.0	&  0.0000  &  7.49$\sigma$  &  -0.337$\pm$0.038  &  10.84$\pm$0.986  &  0.33\\
10.0	&  0.0000  &  8.60$\sigma$  &  -0.355$\pm$0.034  &  11.25$\pm$0.874  &  0.31\\
20.0    &  0.0000  &  7.03$\sigma$  &  -0.293$\pm$0.129  &  9.560$\pm$3.342  &  0.22\\
\enddata

\tablenotetext{a}{Energy at which monochromatic luminosity is taken.}

\tablenotetext{b}{Kendall's probability that correlation is due to chance.}

\tablenotetext{c}{Significance level of Kendall rank correlation coefficient.}
\end{deluxetable}

\begin{figure}
\centering
\includegraphics[width=3.2in]{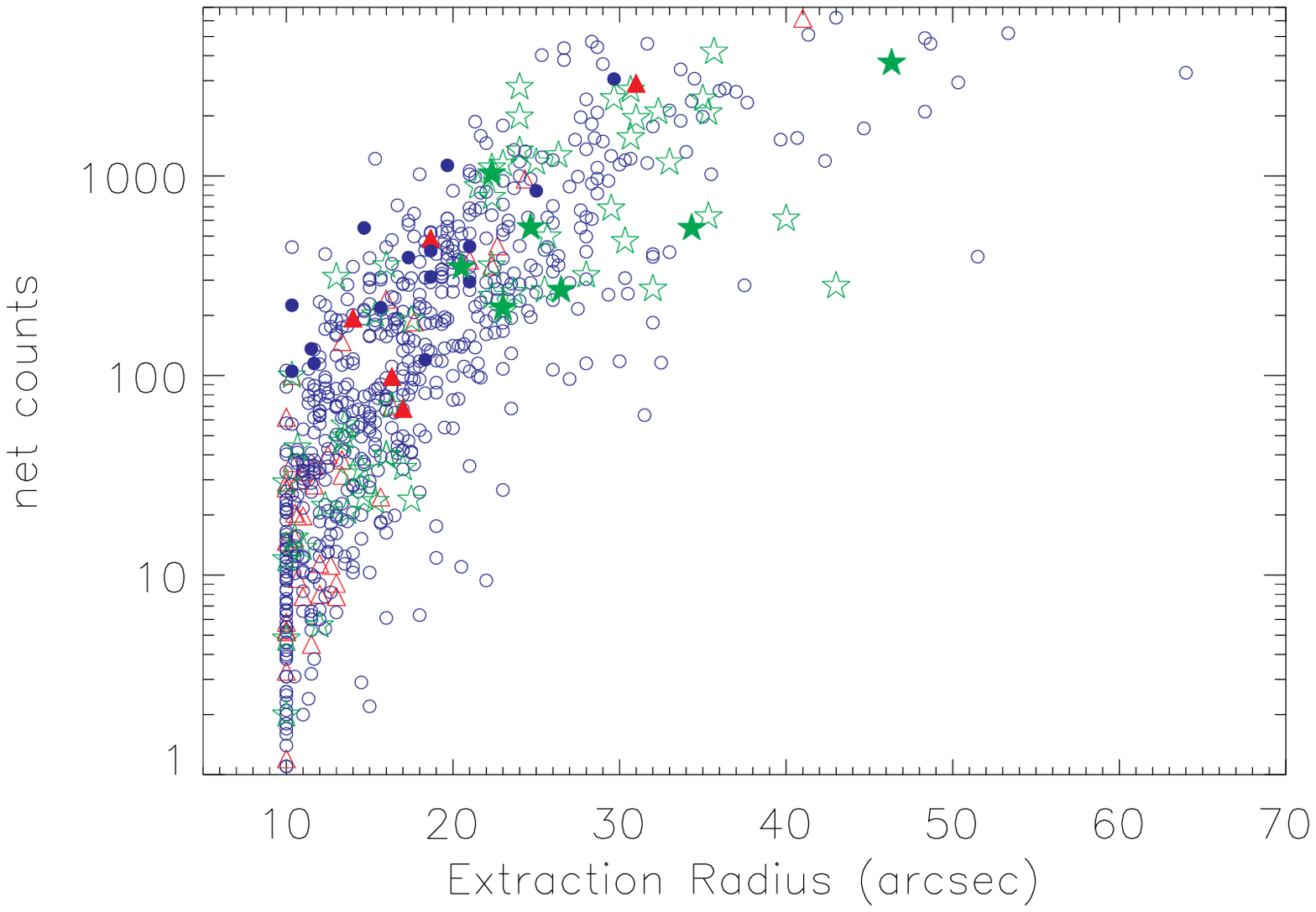}
\includegraphics[width=3.2in]{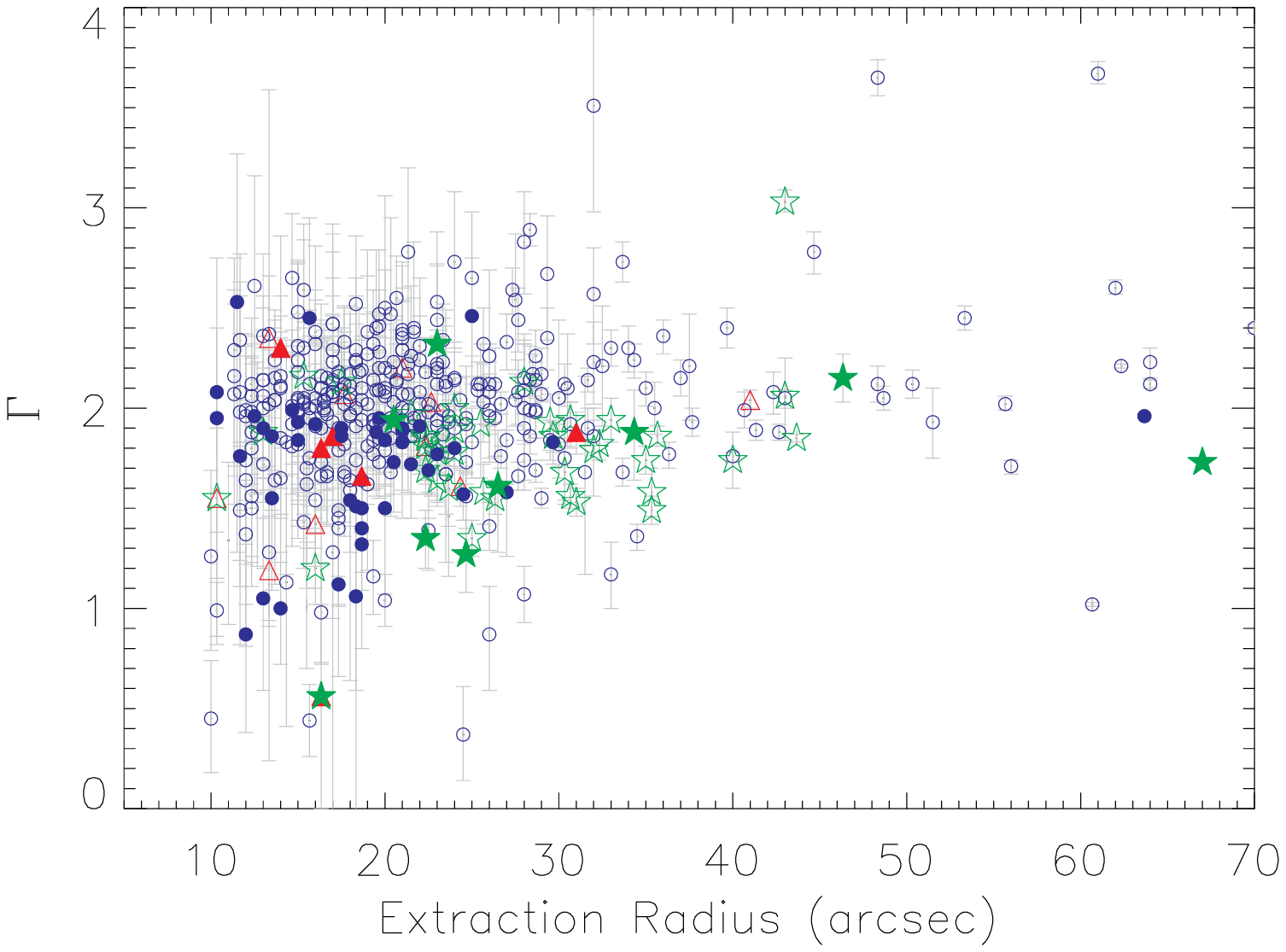}
\includegraphics[width=3.2in]{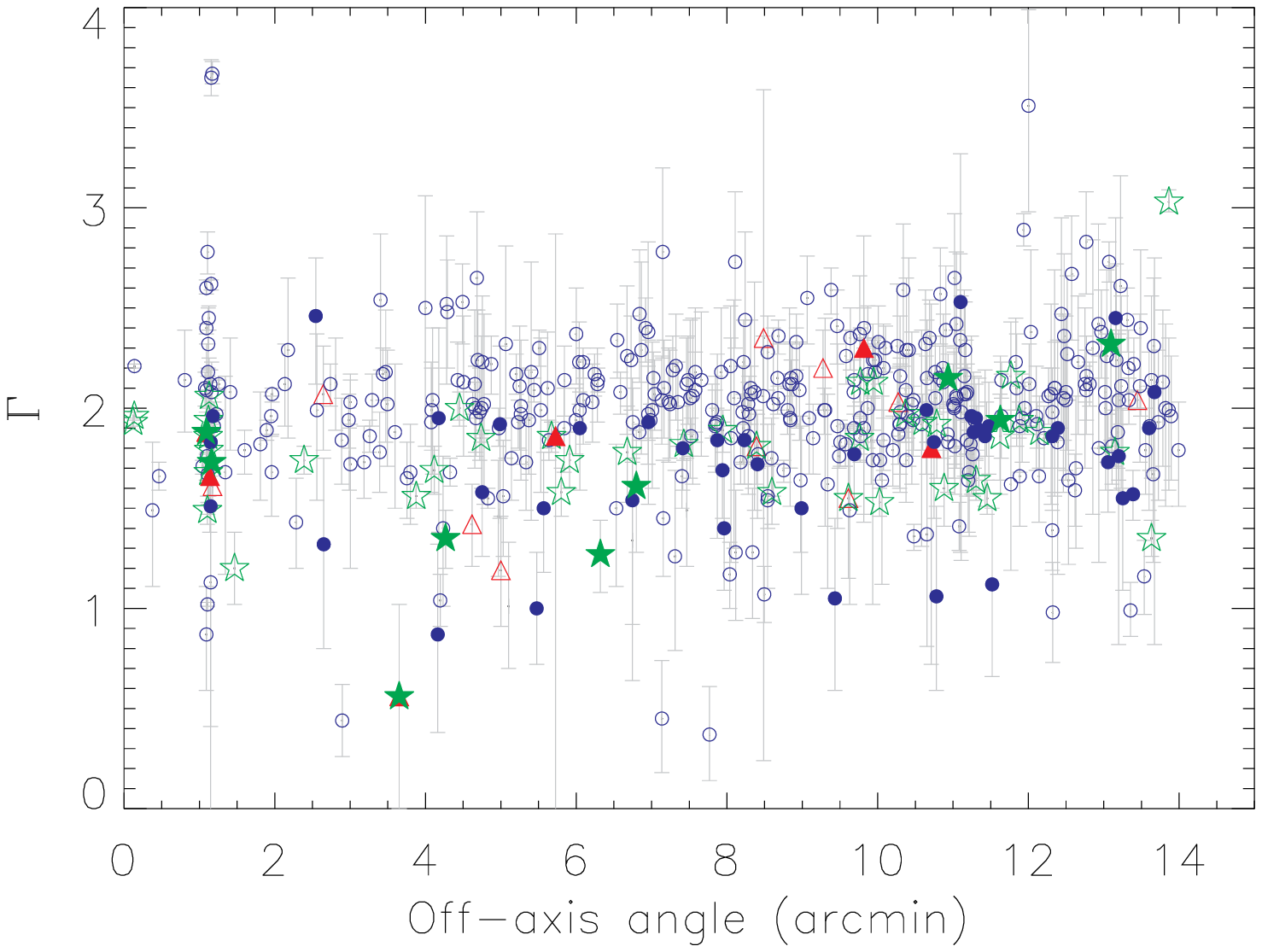}
\includegraphics[width=3.2in]{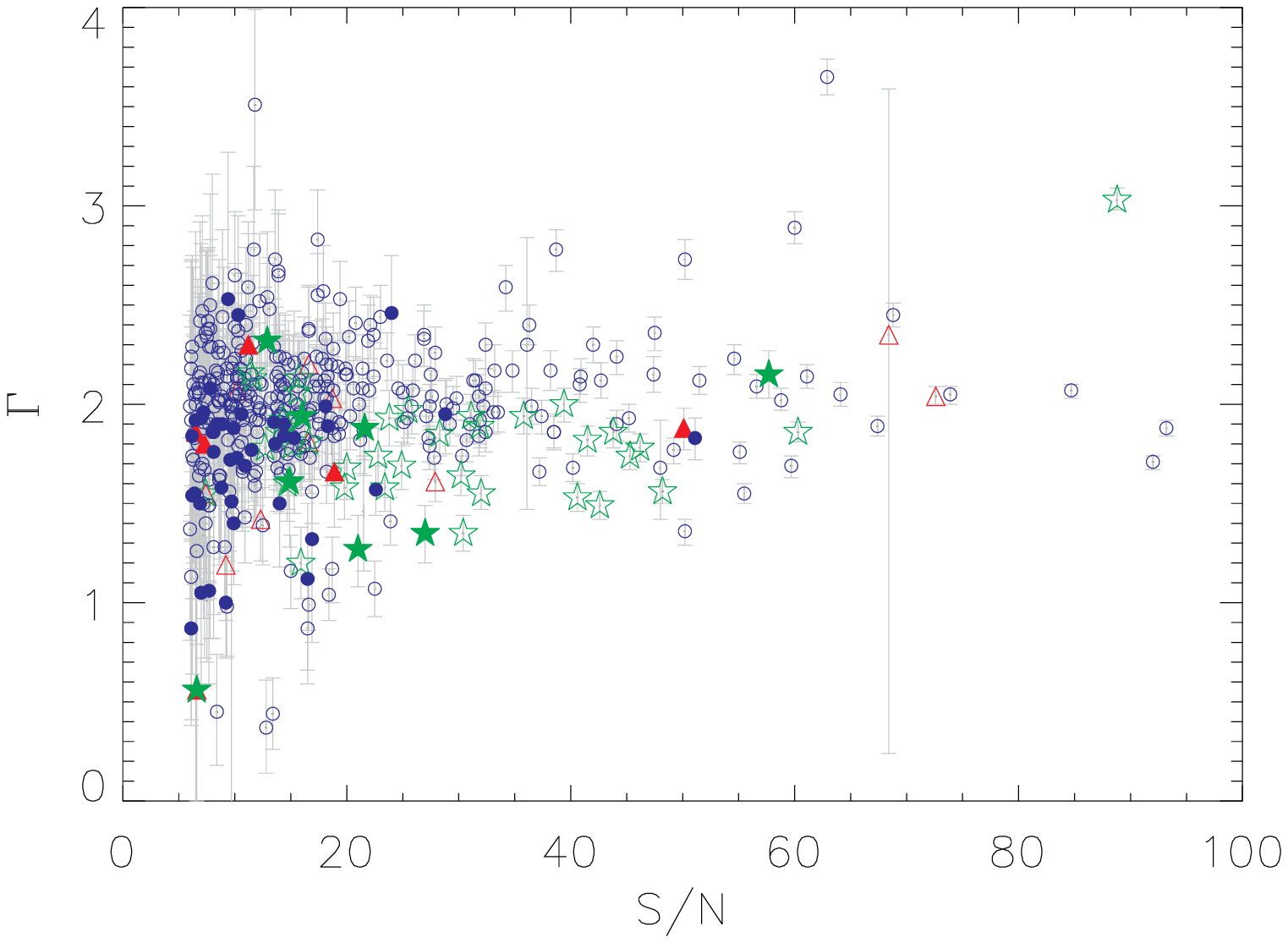}
\caption{(a) Net counts vs. source extraction radius (arcseconds).  (b) $\Gamma$ vs. source extraction radius.  
(c) X-ray photon index ($\Gamma$) vs. off-axis angle (arcmin).  (d) $\Gamma$ vs. S/N. }
\label{fig:extraction}
\end{figure}

\begin{figure}
\centering
\includegraphics[width=3.2in]{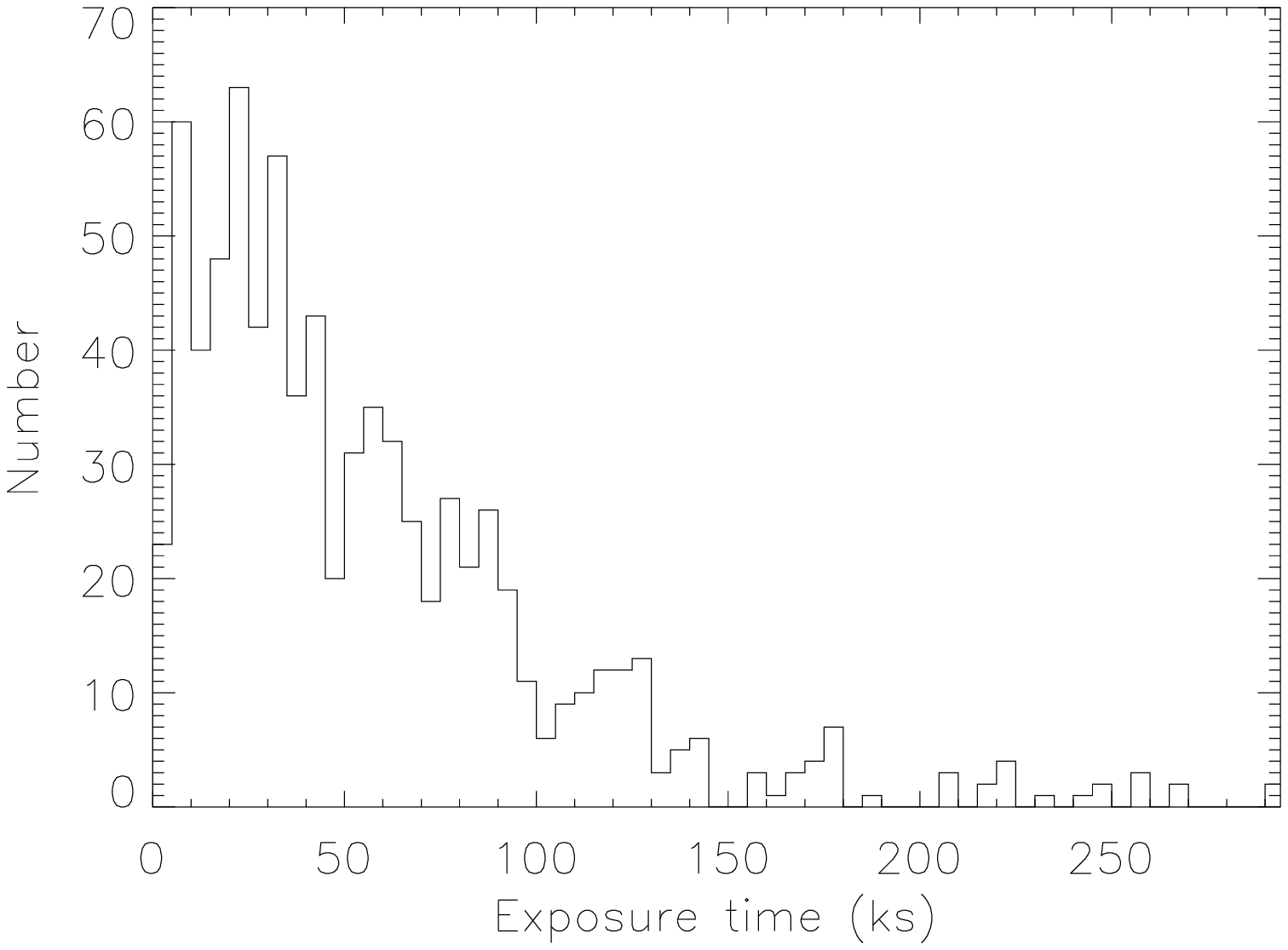}
\includegraphics[width=3.2in]{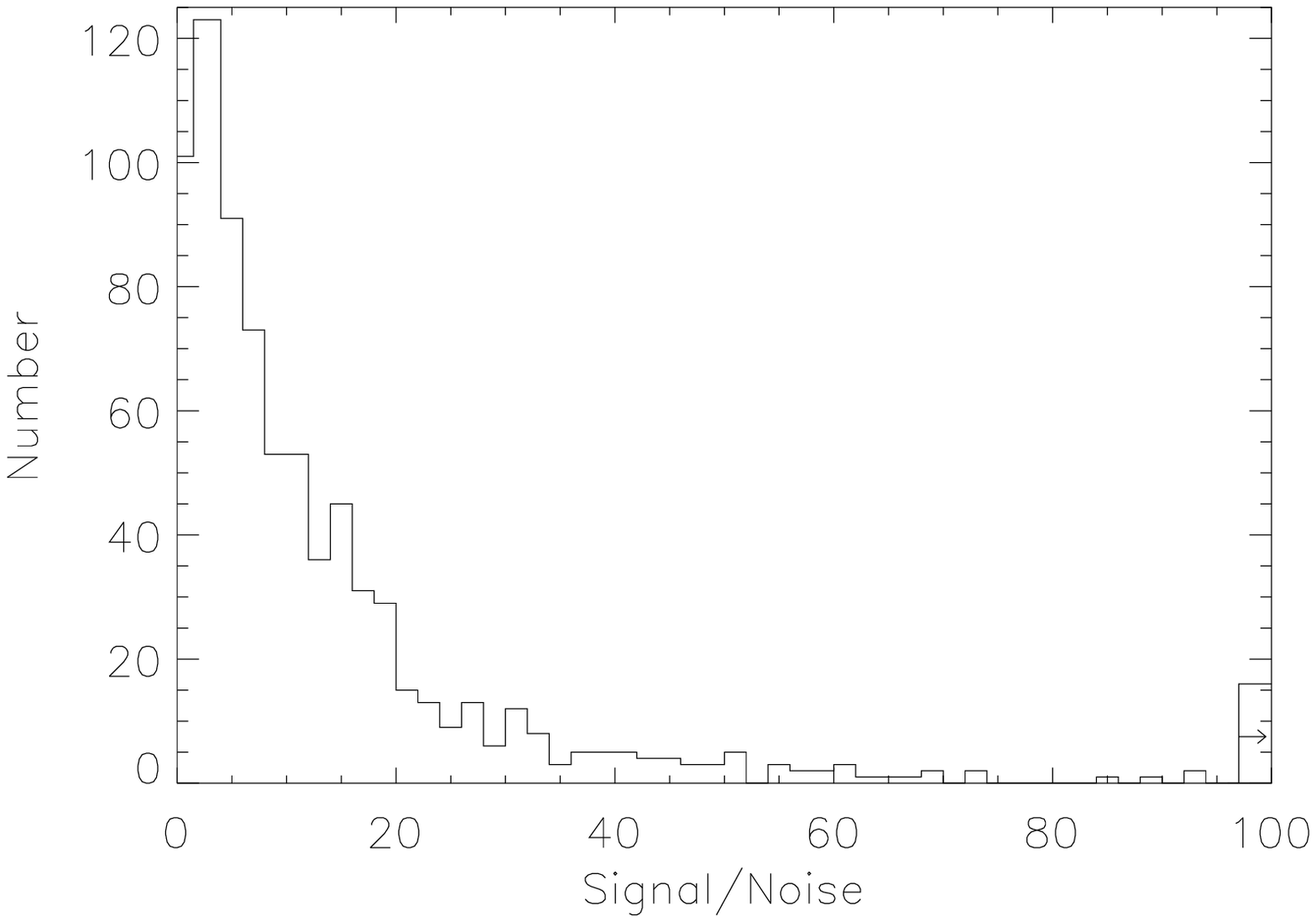}
\includegraphics[width=3.2in]{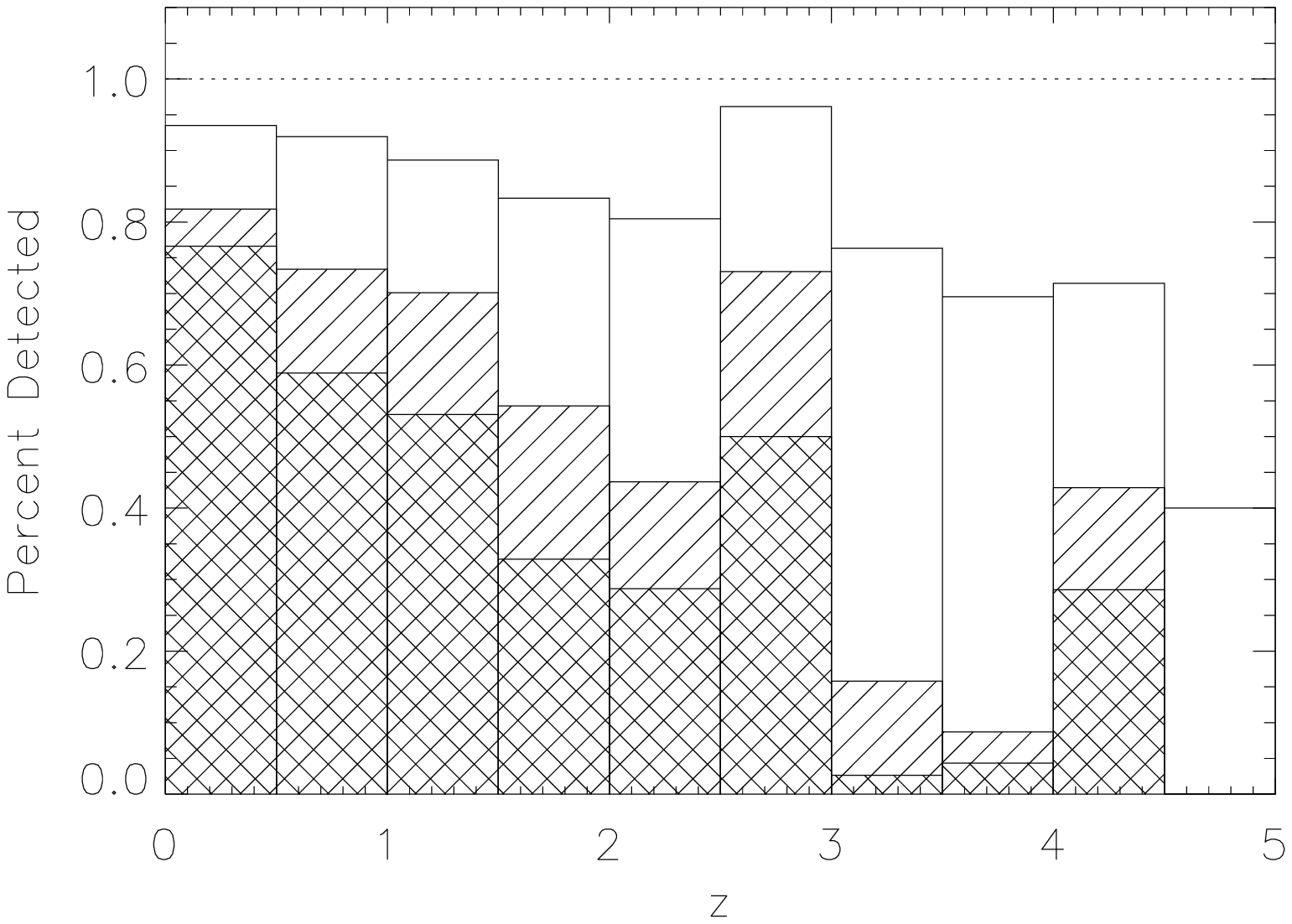}
\includegraphics[width=3.2in]{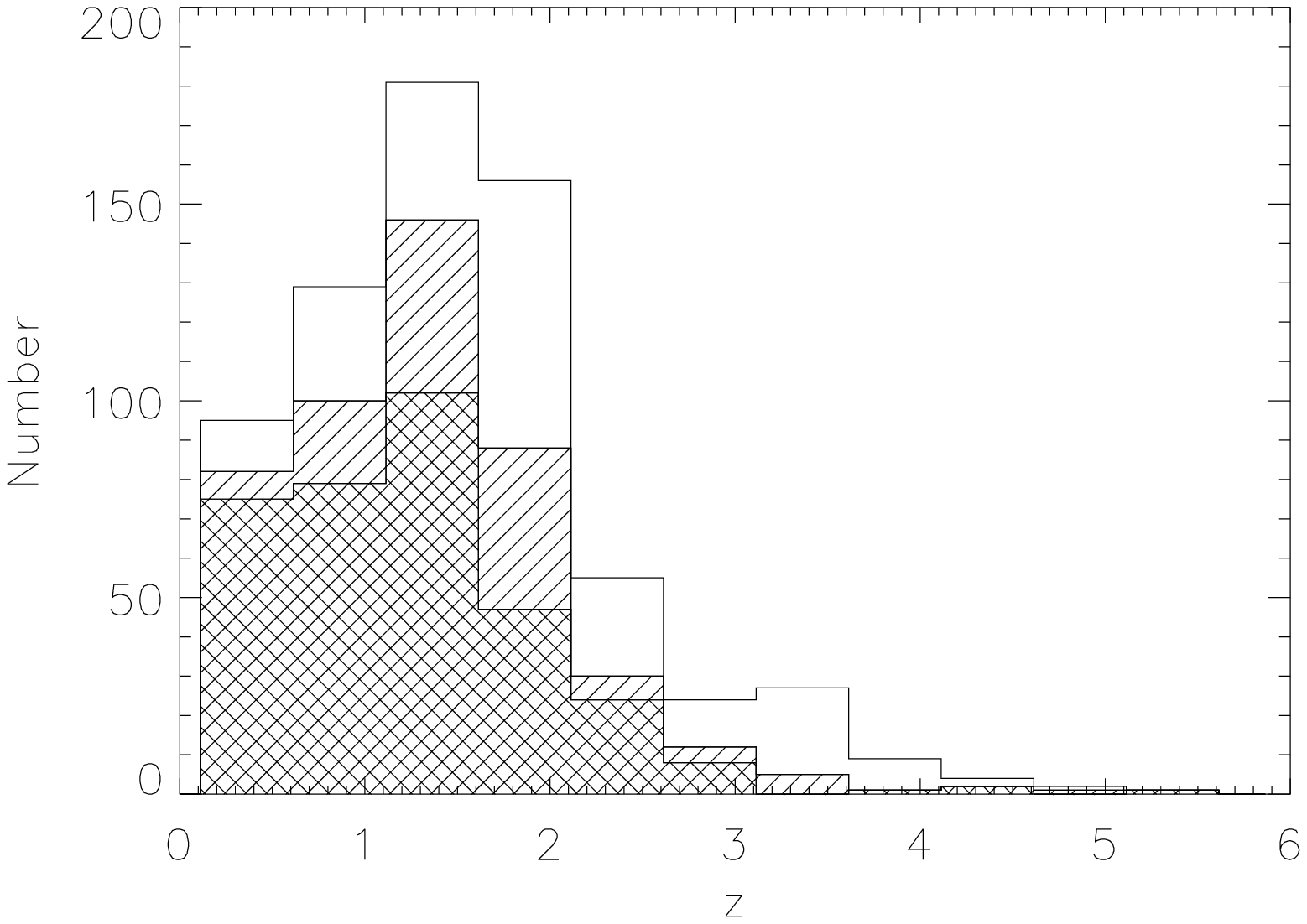}
\caption{Survey characteristics: (a) Exposure time histogram, where exposure times are summed over the up 
to three EPIC cameras.  Exposure times do not include high-background intervals filtered out during data 
reduction.  (b) S/N histogram, where 16 sources have X-ray S/N $>$ 100.  (c) Fraction of detected sources 
vs. redshift, and (d) redshift histogram.  For (c) and (d), the open, dotted-line histogram is for all 
SDSS-selected quasars, the open, solid-line histogram is for all detected sources, the hatched histogram 
is for sources with sources with X-ray S/N $>$ 6, and the double-hatched histogram is for sources with 
X-ray S/N $>$ 10.}
\label{fig:characteristics}
\end{figure}

\begin{figure}
\centering
\includegraphics[width=3.2in]{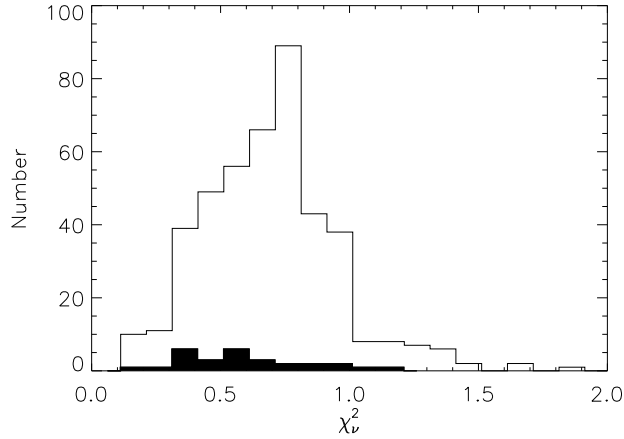}
\caption{$\chi^2_\nu$ histogram.  The open histogram represents sources that prefer a power-law model.  The 
solid histogram represents sources that prefer an absorbed power-law model.  Five sources have $\chi^2_\nu$ 
$>$ 2 and are not included in the plot.}
\label{fig:chi2}
\end{figure}

\begin{figure}
\centering
\includegraphics[width=3.2in]{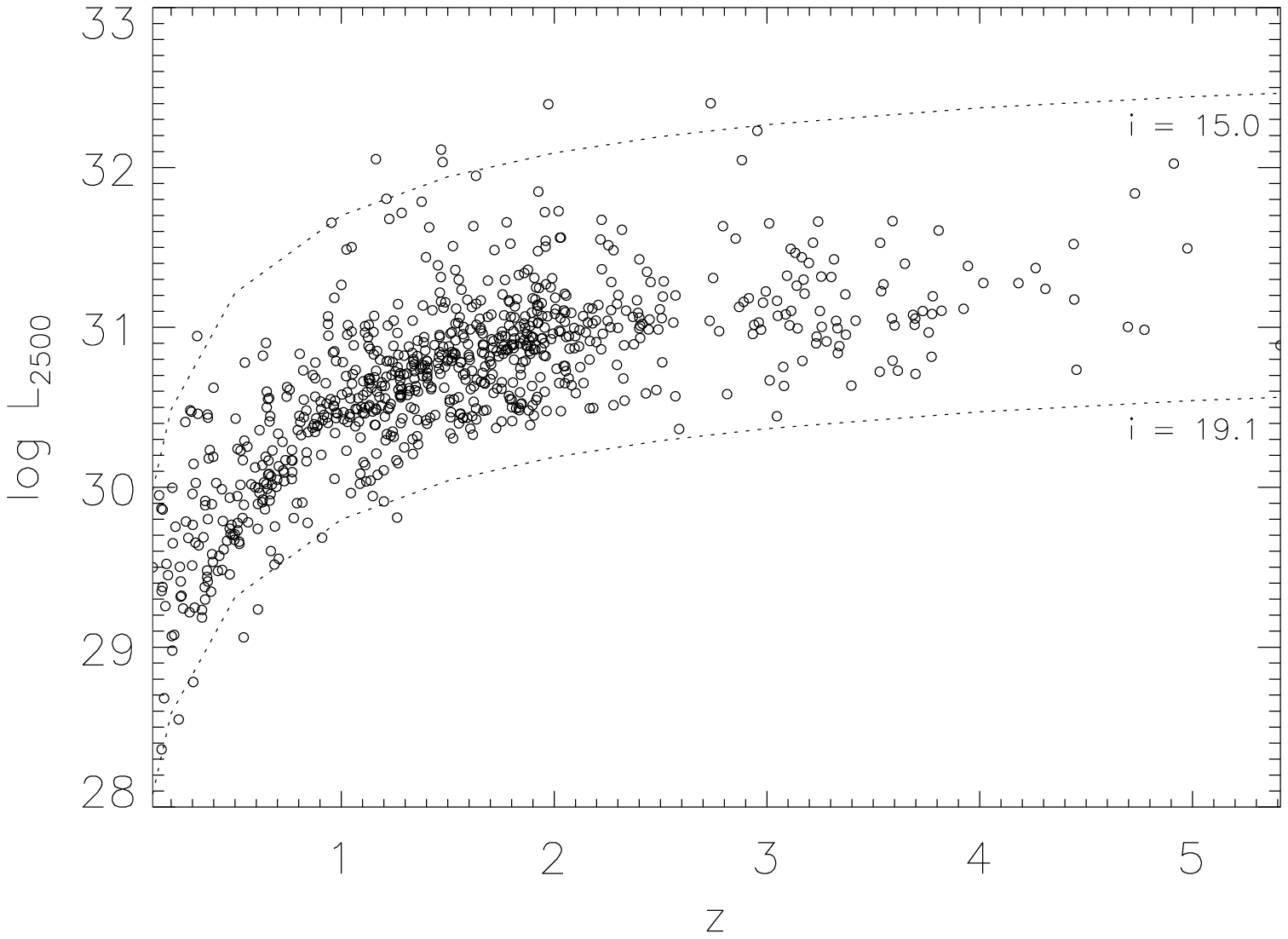}
\includegraphics[width=3.2in]{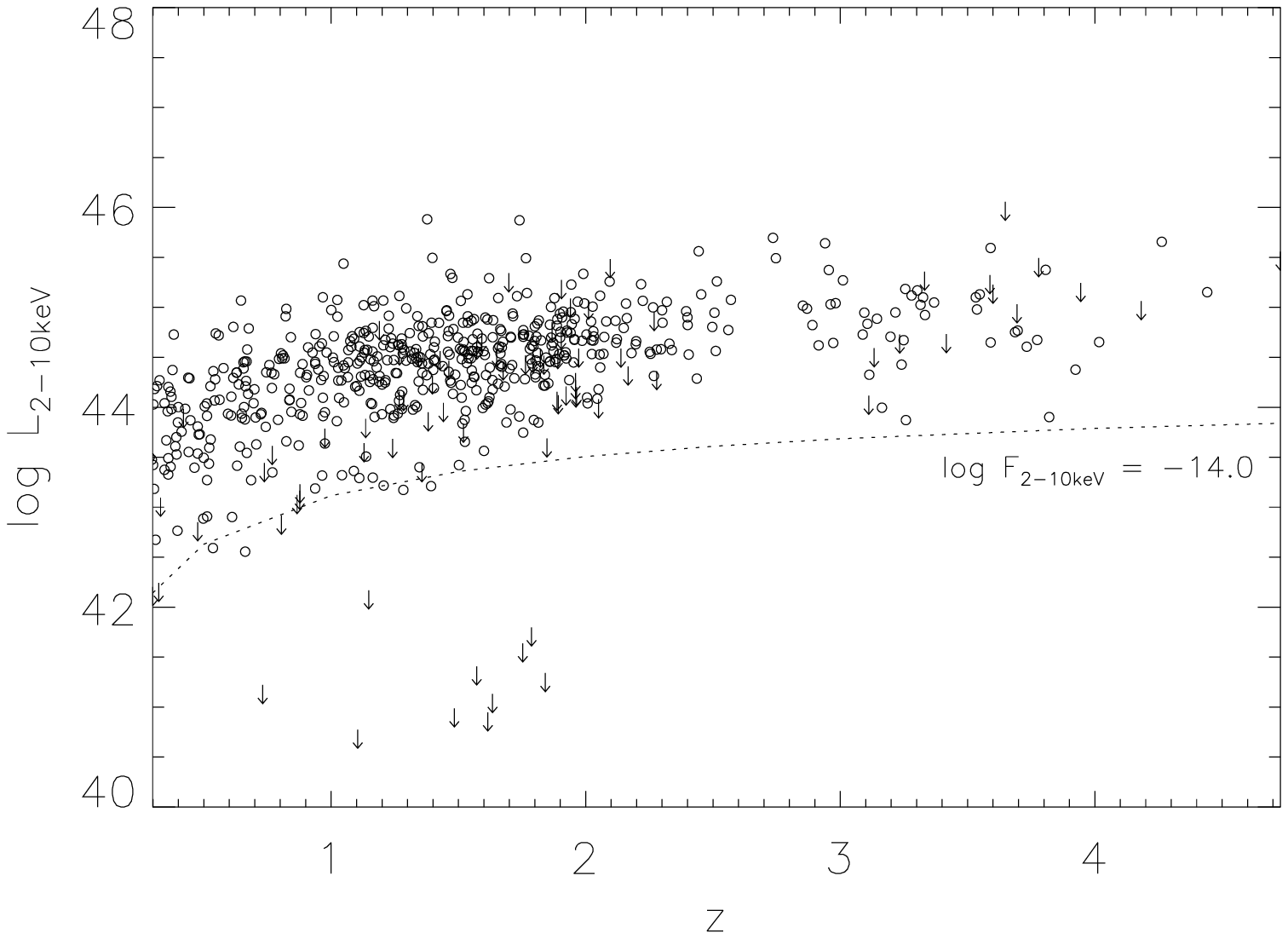}
\includegraphics[width=3.2in]{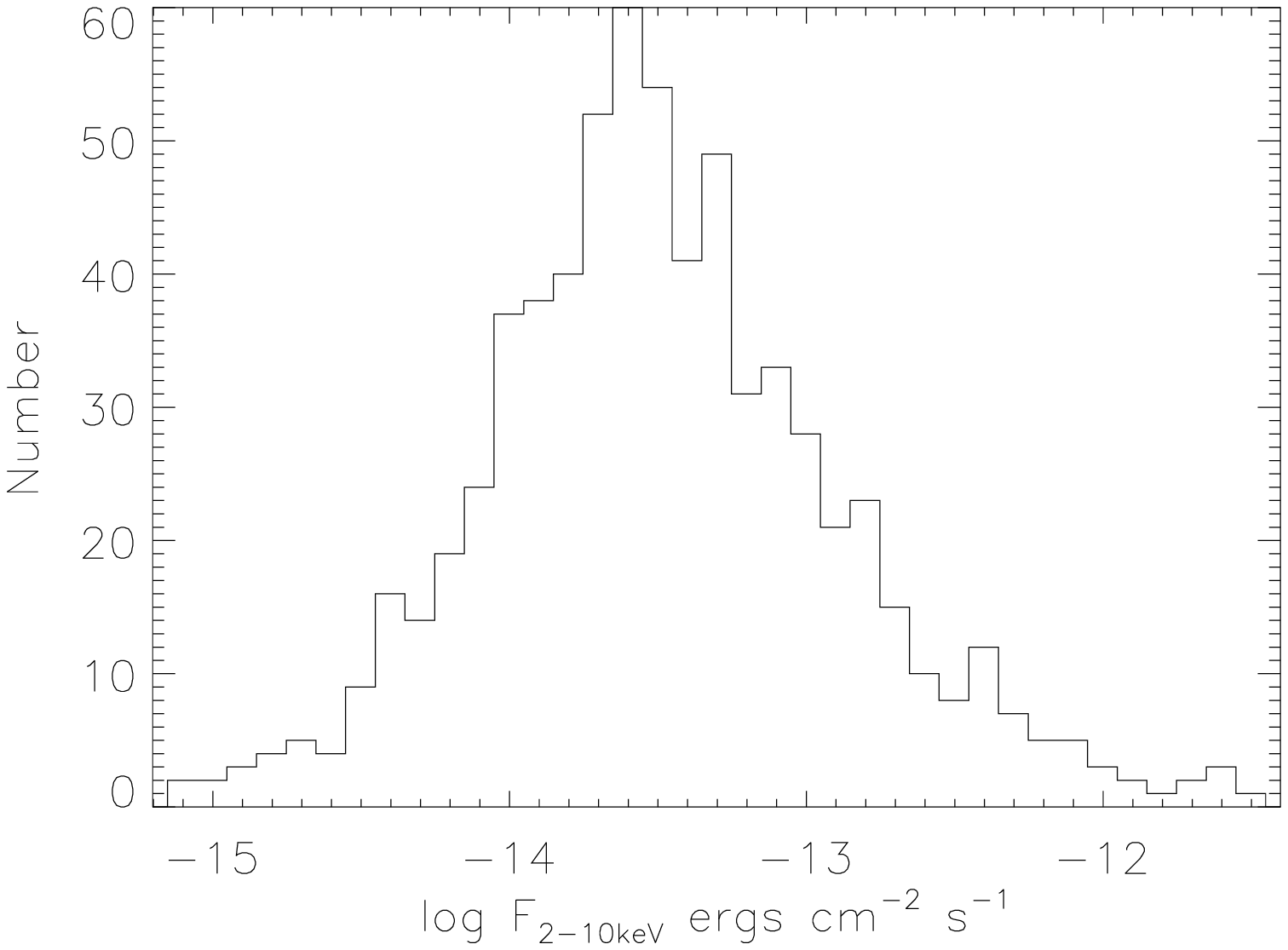}
\caption{X-ray and optical sensitivity: (a)  Optical 2500 $\mbox{\AA}$ luminosity vs. redshift and 
(b) X-ray 2-10 keV luminosity vs. redshift.  Flux limits are plotted in both plots.  For (a), $i$ 
band magnitude limits (dotted lines) were set during selection \citep{Richards02}.  For (b), the 
flux limit (log F$_{2-10keV}$ = -14.0 ergs cm$^{-2}$ s$^{-1}$, dotted line) is for an observation 
of 20 ks \citep{Watson01}.  (c) The observed-frame F$_{2-10keV}$ distribution.}
\label{fig:sensitivity}
\end{figure}

\begin{figure}
\centering
\includegraphics[width=3.2in]{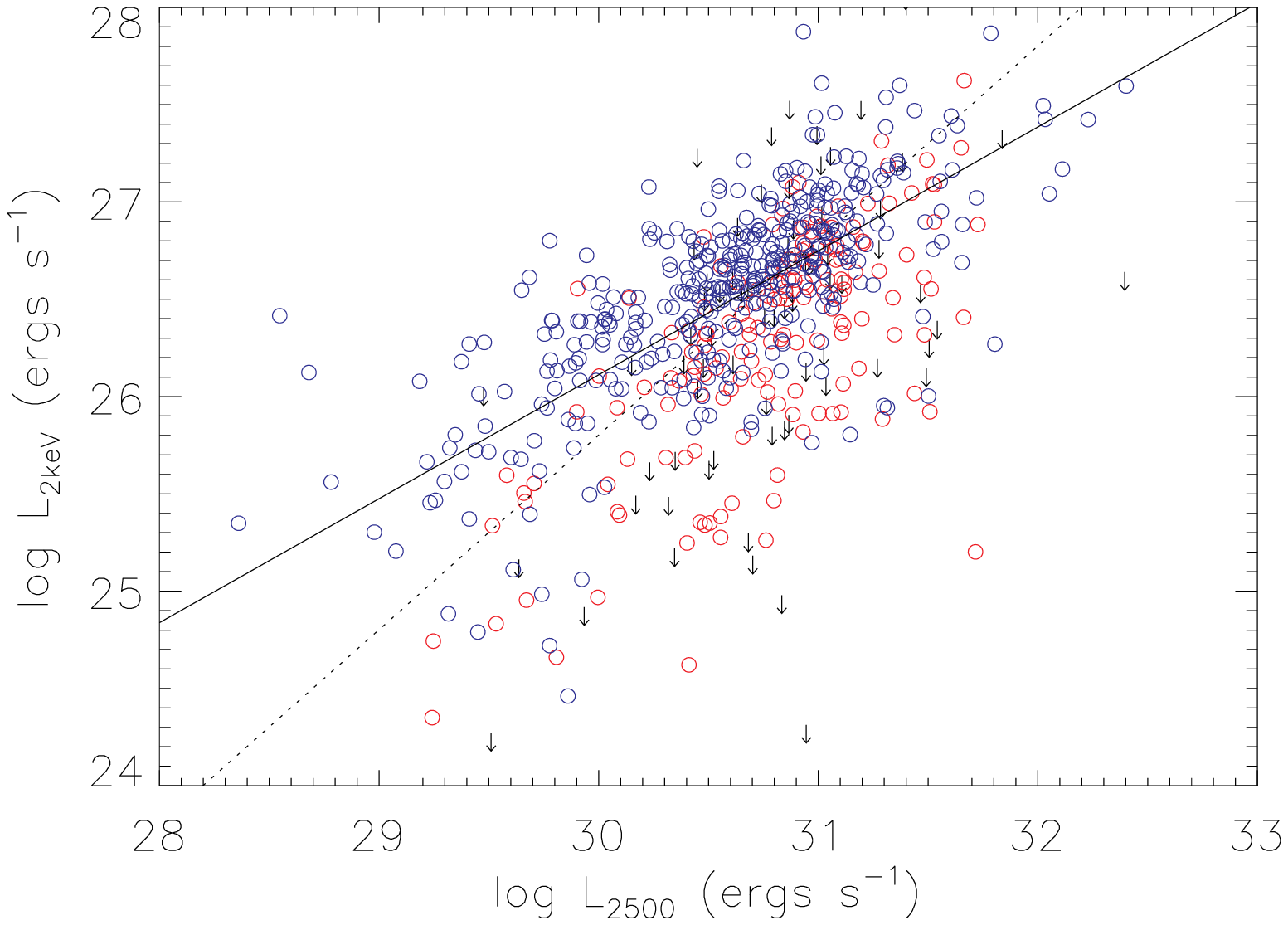}
\includegraphics[width=3.2in]{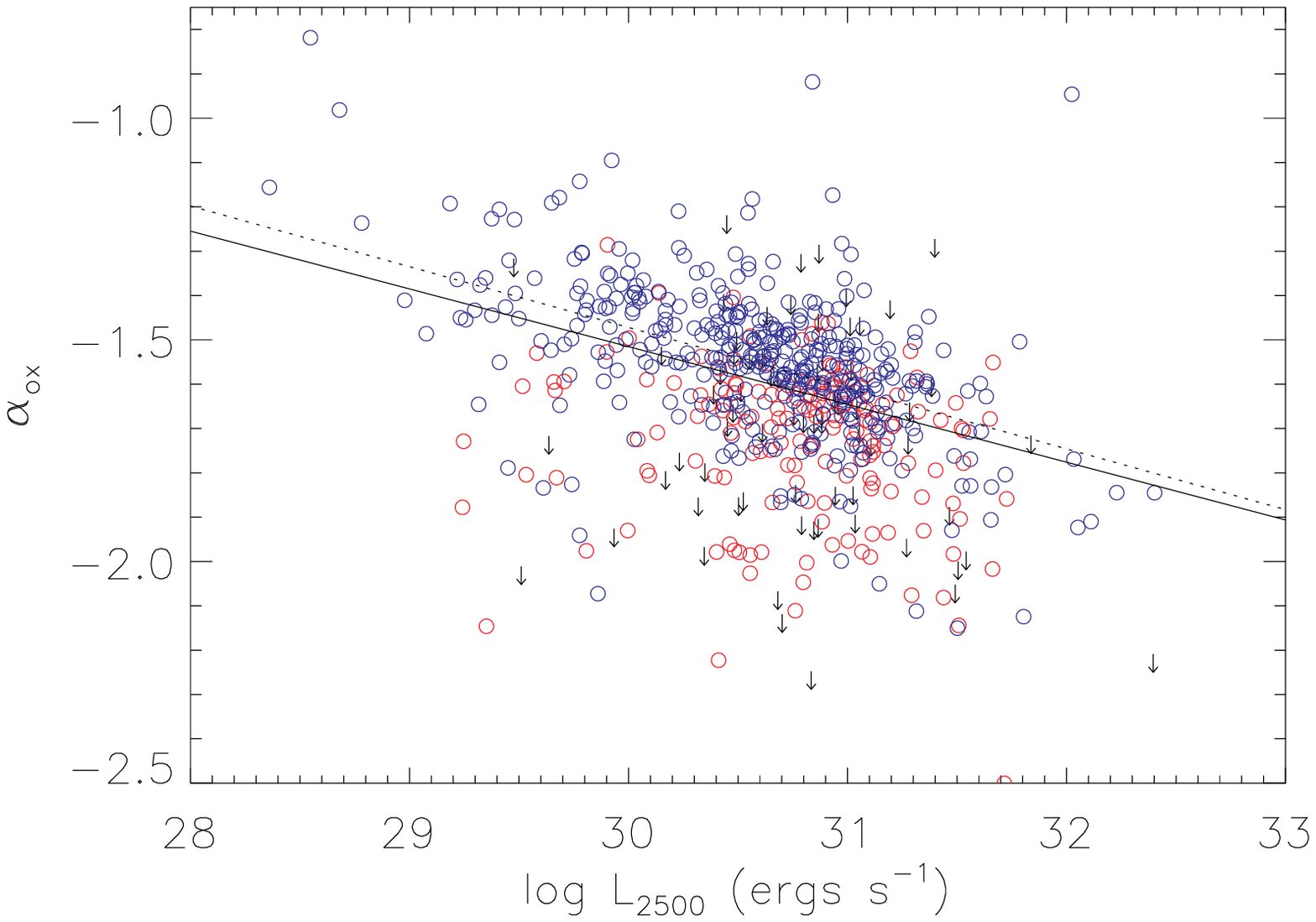}
\includegraphics[width=3.2in]{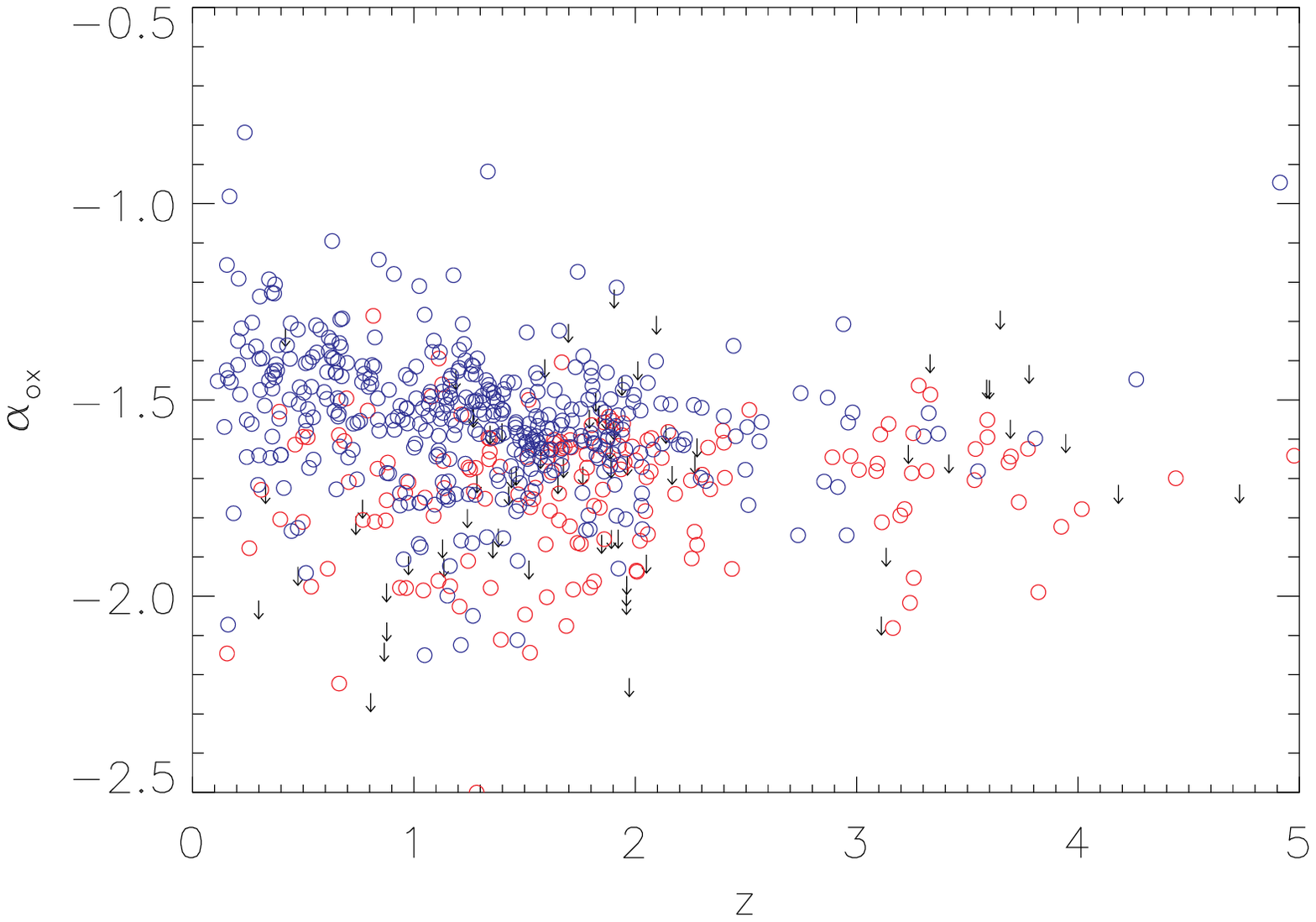}
\caption{(a) $l_{2keV}$ vs. $l_{2500}$.  The dotted line shows $l_{2keV} \propto l_{2500}$, 
constrained to pass through the average values of $l_{2keV}$ and $l_{2500}$.  (b) The $\alpha_{ox}-l_{2500}$ 
relation for the SDSS/XMM quasar sample.  The linear fit to the SDSS/XMM sample is plotted as a solid line, 
and the Steffen et al. line is overplotted as a dotted line for comparison.  (c) $\alpha_{ox}$ vs. redshift.  
Red circles represent detected sources with X-ray S/N $<$ 6, and blue circles represent sources 
with X-ray S/N $>$ 6.  BALs and RL quasars were removed from all three plots.}
\label{fig:aox}
\end{figure}

\begin{figure}
\centering
\includegraphics[width=3.2in]{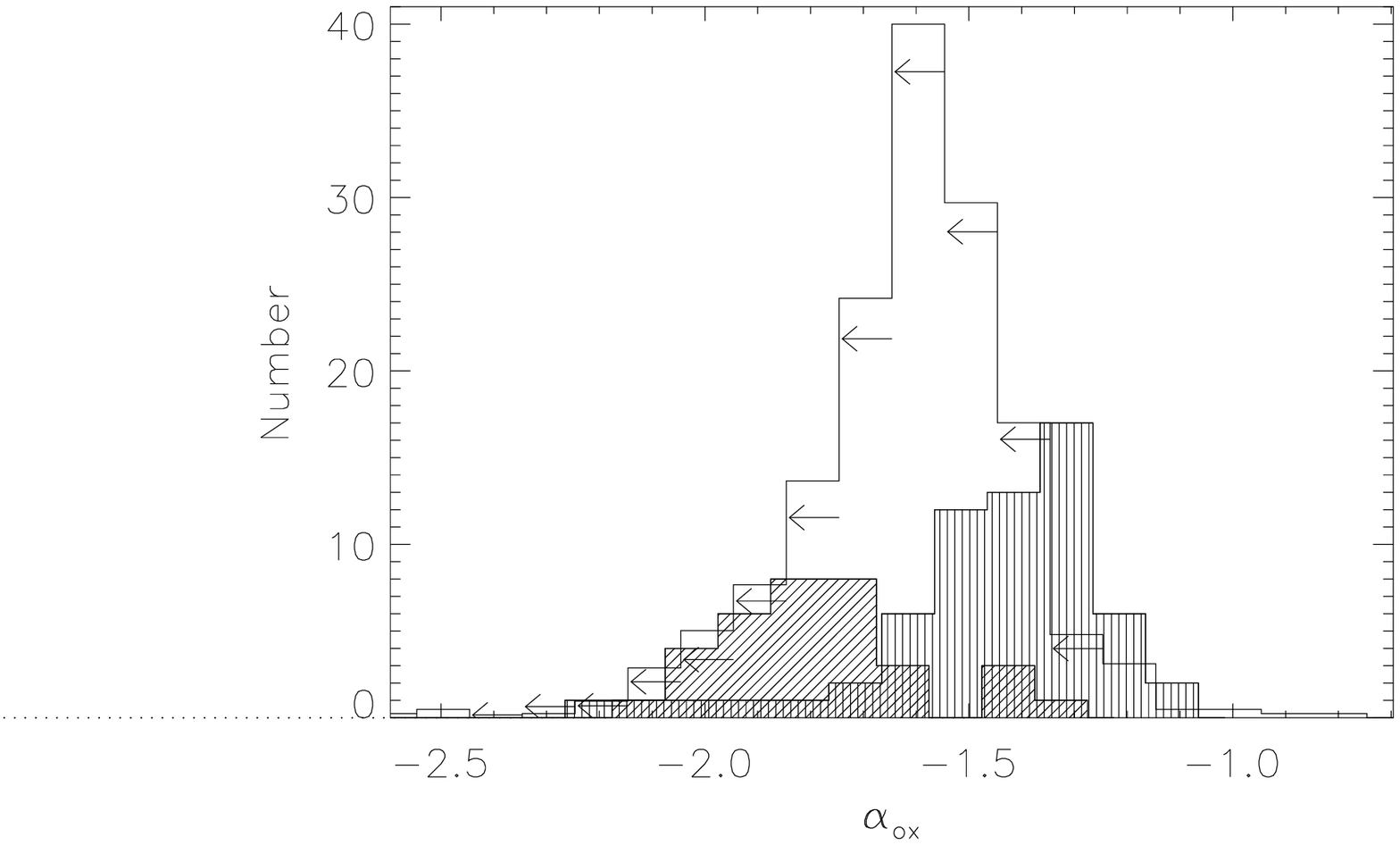}
\includegraphics[width=3.2in]{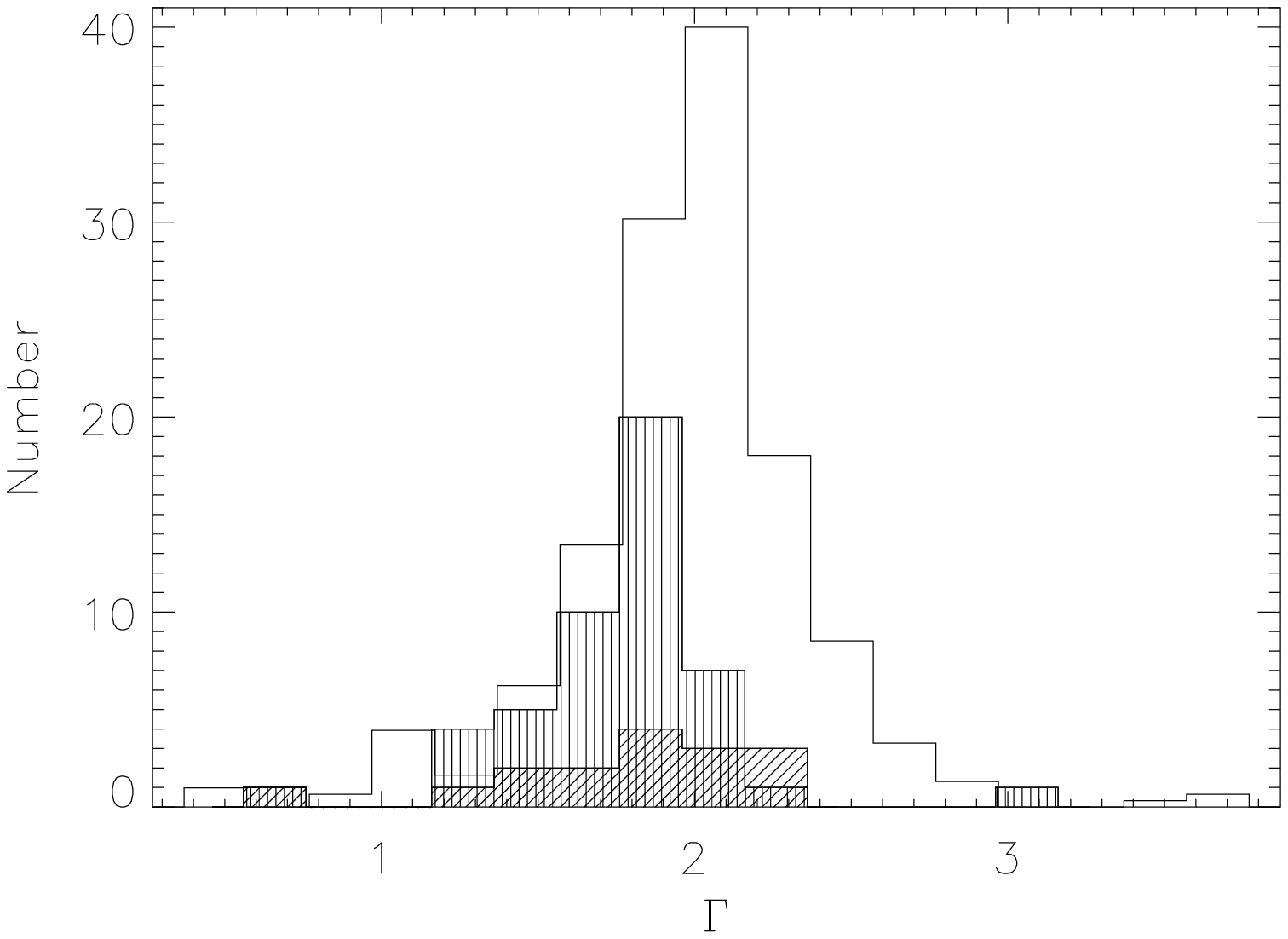}
\caption{(a) $\alpha_{ox}$ distribution for all sources.  (b) $\Gamma$ distribution for sources with 
S/N $>$ 6.  Open histograms represent RQ, non-BAL quasars and have been scaled down by factors of 3 and 4, 
respectively.  Histograms with vertical hatching represent RL quasars, and histograms with diagonal 
hatching represent BAL quasars.  The space above the upper-limit arrows indicates the number of RQ+non-BAL 
upper-limits in a given bar.}
\label{fig:HistaoxGamma}
\end{figure}

\begin{figure}
\centering
\includegraphics[width=3.2in]{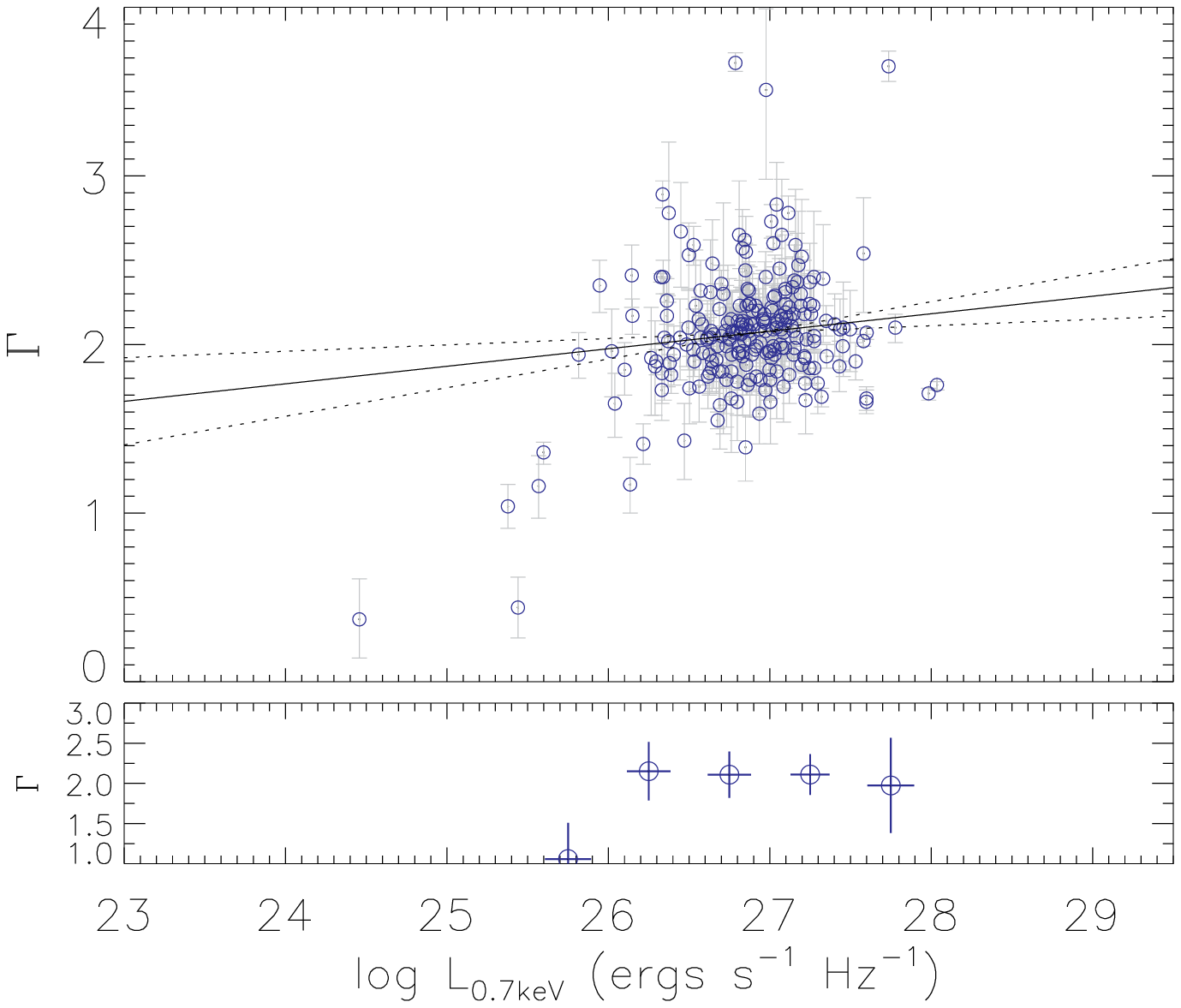}
\includegraphics[width=3.2in]{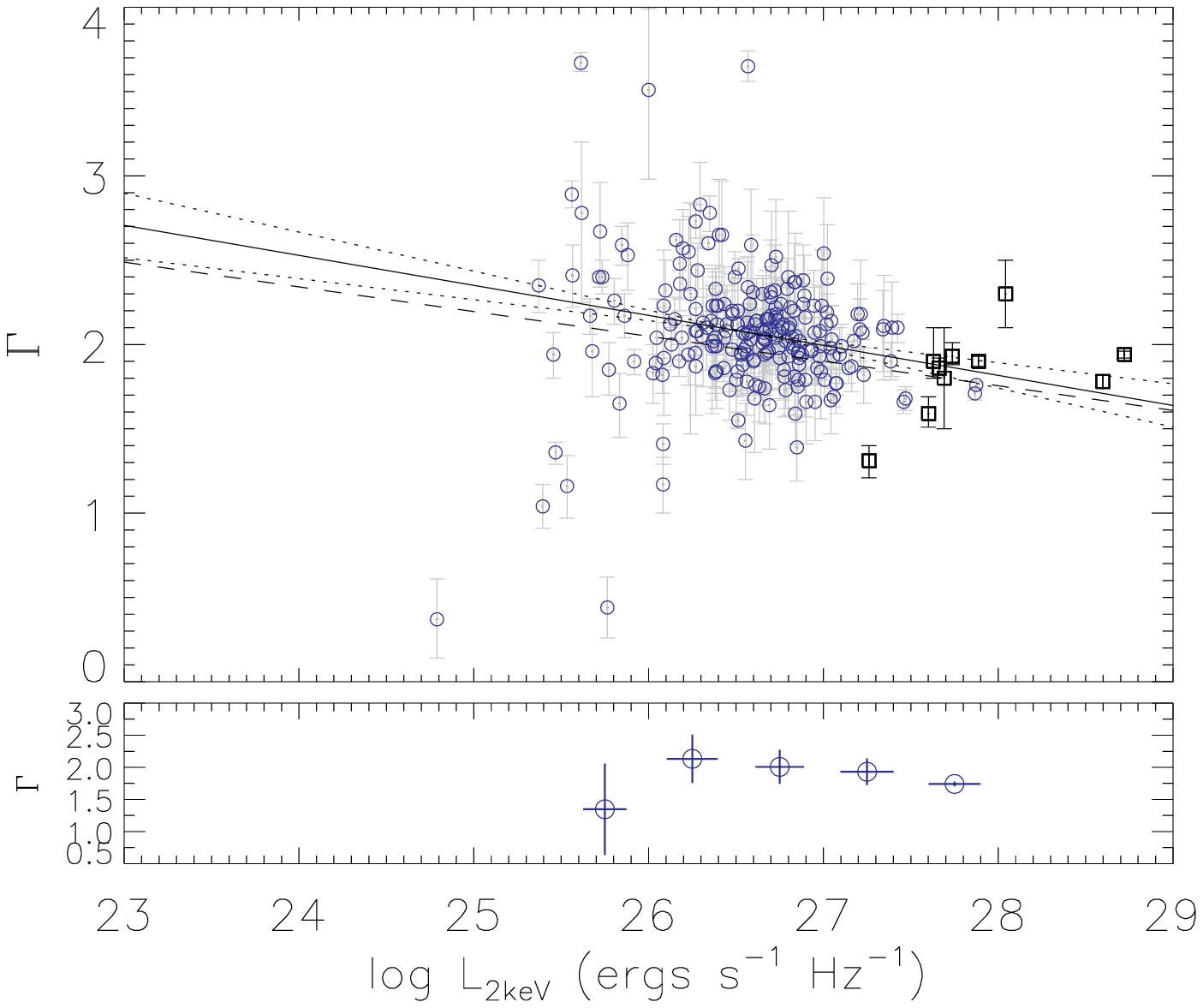}
\includegraphics[width=3.2in]{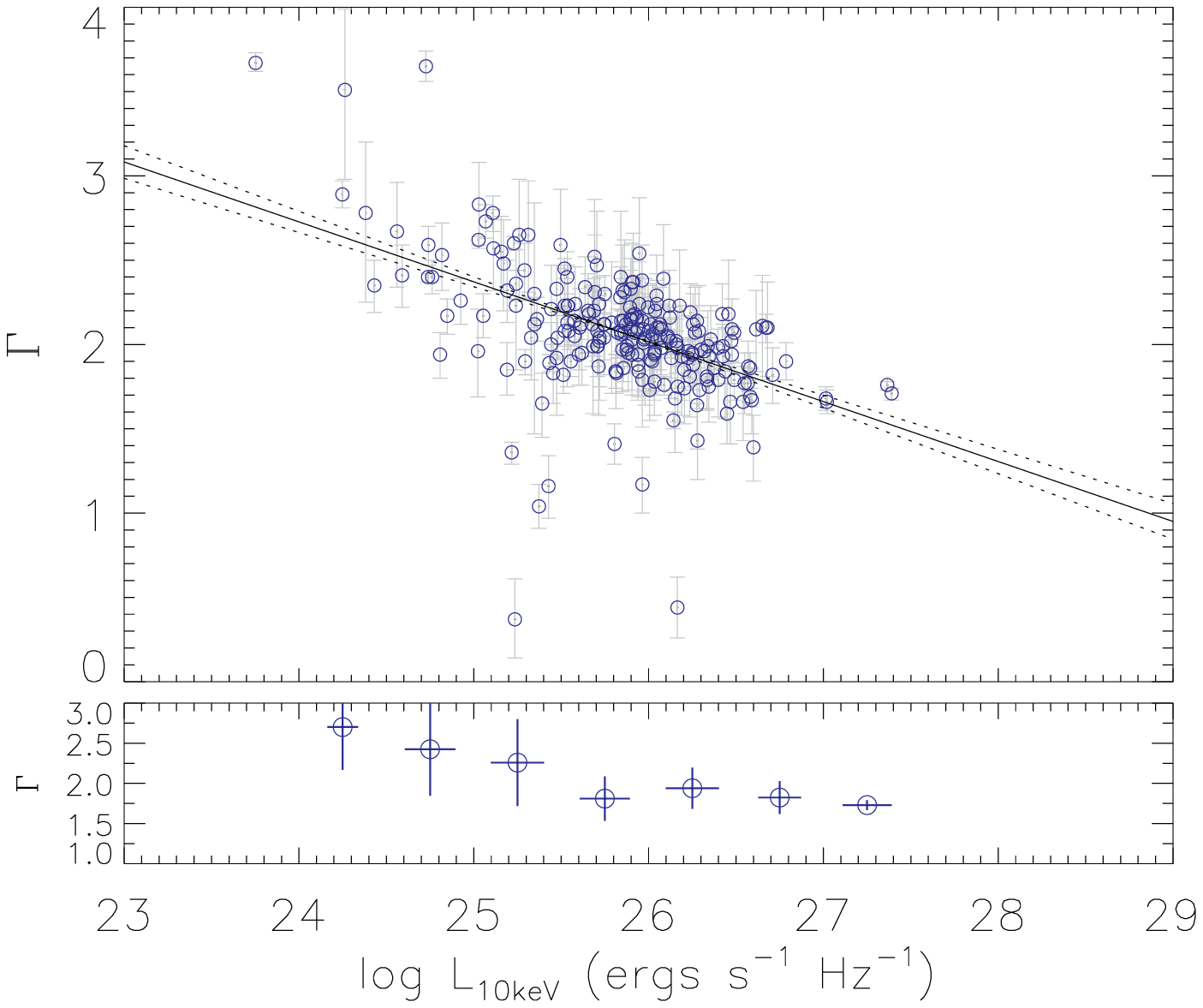}
\includegraphics[width=3.2in]{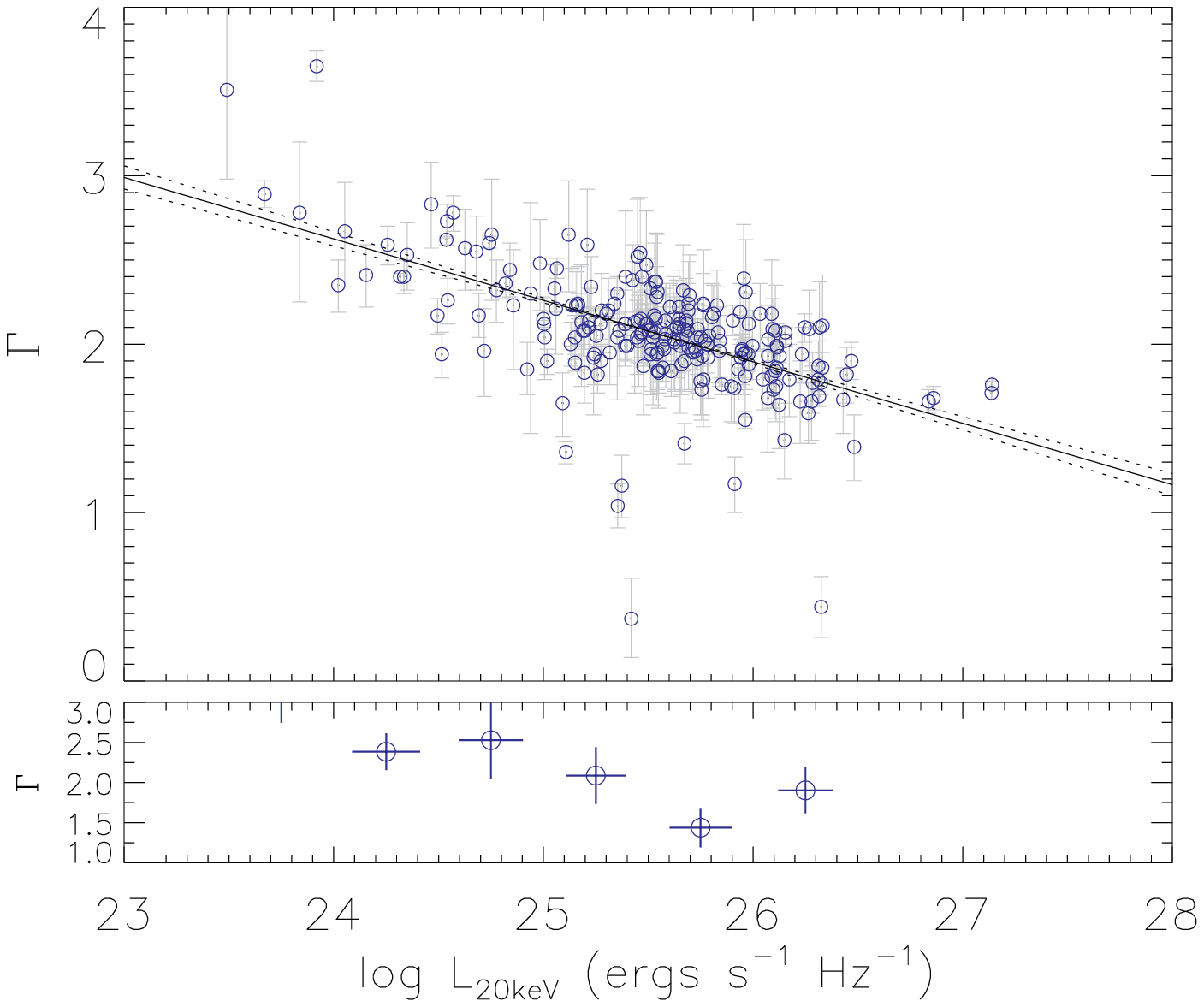}
\caption{TOP:  X-ray photon index ($\Gamma$) vs. monochromatic X-ray luminosity at rest-frame (a) 0.7 keV, 
(b) 2 keV, (c) 10 keV, and (d) 20 keV.  In (b), the \citet{Green08} correlation is plotted for comparison 
as a dashed line, and the quasars from \citet{Dai04} are over-plotted as open black squares.  
Only RQ, non-BAL quasars are plotted \emph{(blue circles)}
The weighted least squares regression line is calculated using RQ, non-BAL quasars, and is 
plotted as a solid line, with errors shown as dotted lines.  
BOTTOM: The weighted mean $\Gamma$ values are given for bins of width $\Delta$log L$_{2-10keV}$ = 0.5.  
Error lines mark the 1$\sigma$ dispersions in both axes.  Blue lines represent RQ+non-BAL quasars, while 
green lines represent RL+non-BAL quasars.}
\label{fig:GammaLx}
\end{figure}

\begin{figure}
\centering
\includegraphics[width=6in]{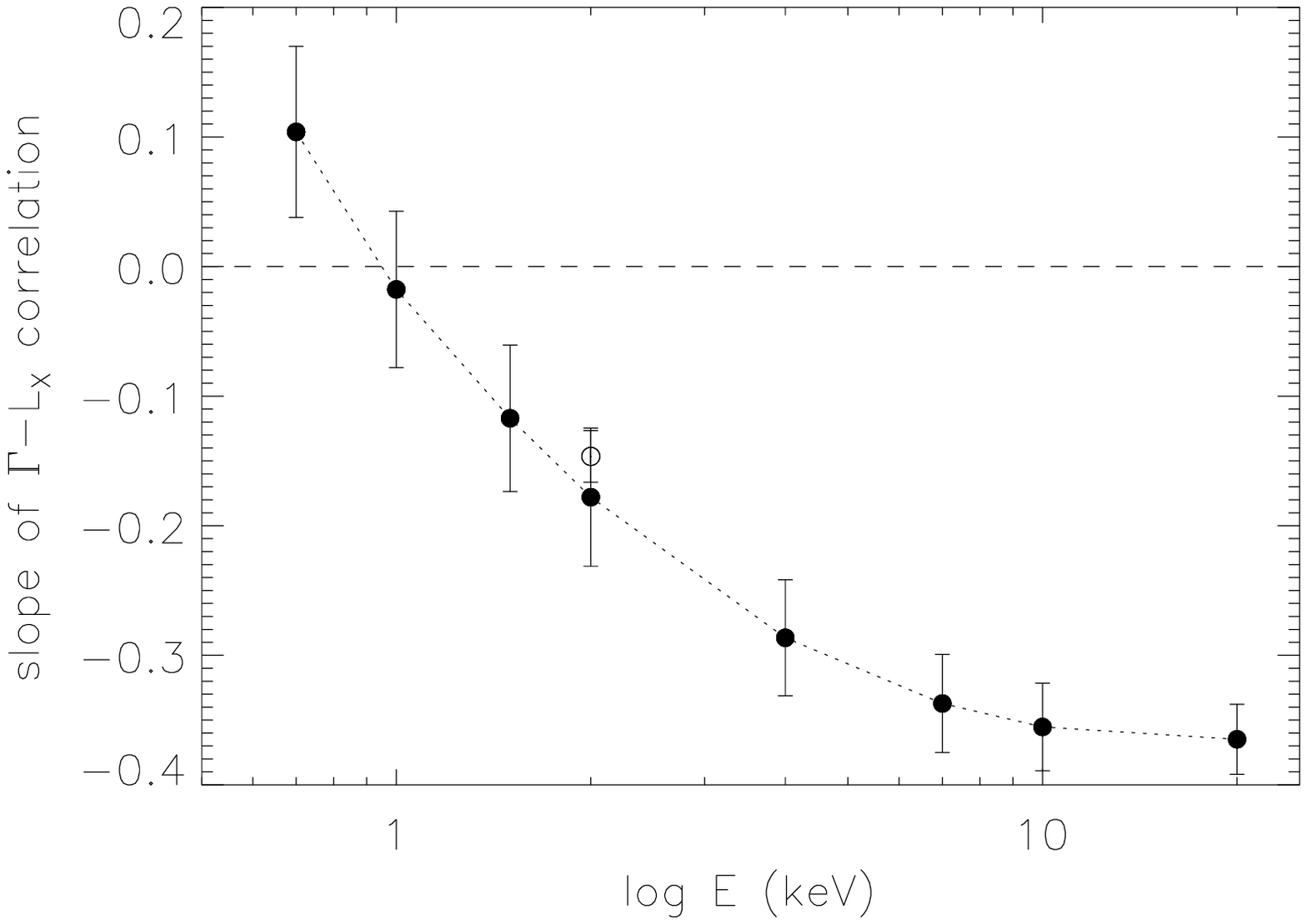}
\includegraphics[width=6in]{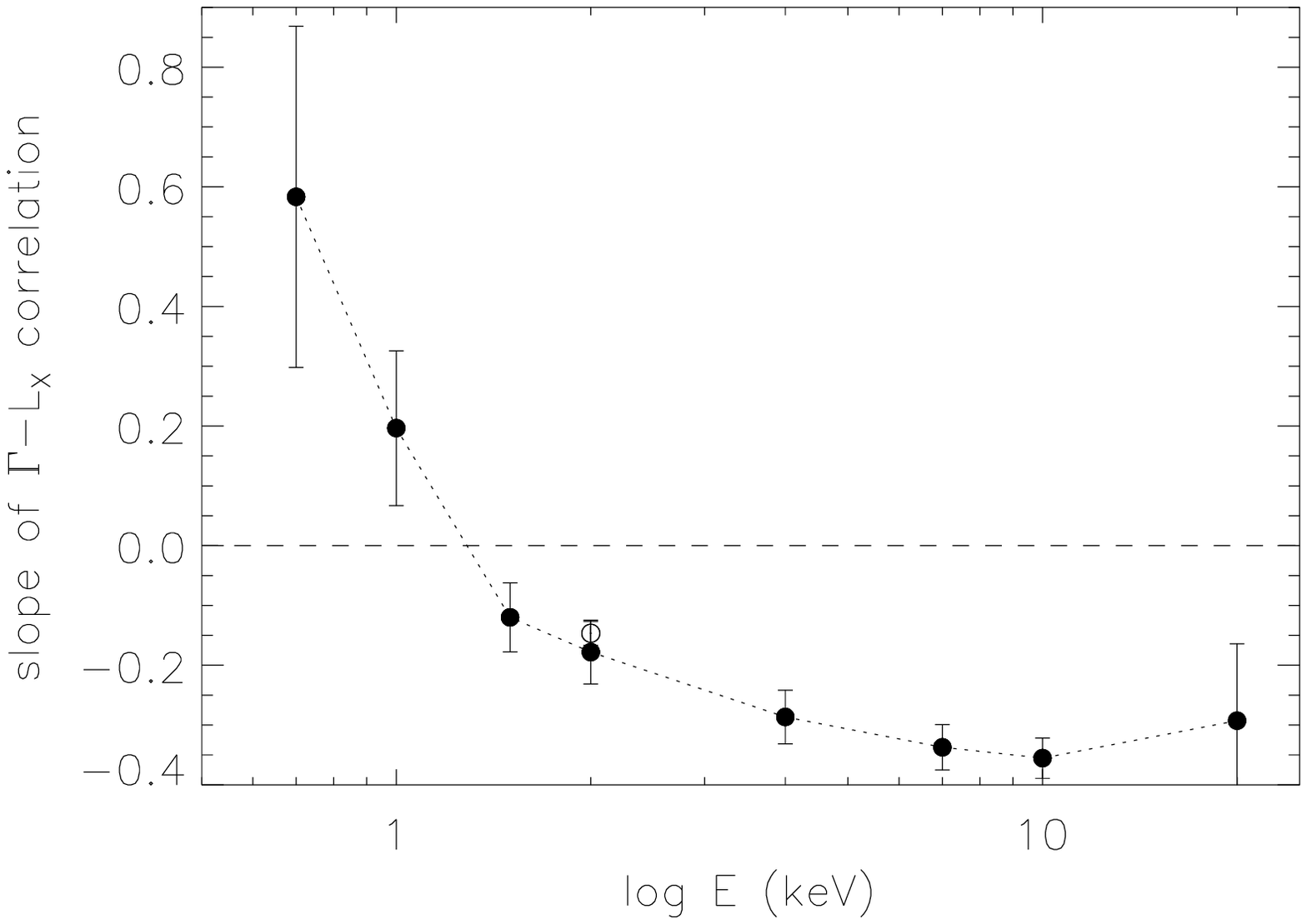}
\caption{The slope of the $\Gamma$-L$_X$ correlation vs. energy E with 1$\sigma$ errors on the slope 
for (a) all sources and (b) sources where the rest-frame energy lies in the observed range.  
The slopes of the $\Gamma$-L$_X$ correlations are obtained for E = 0.7, 1, 1.5, 2, 4, 7, 10, and 20 keV, 
and are marked as solid circles.  
The \citet{Green08} correlation slope is marked as an open circle at 2 keV.  A dashed line marks zero.}
\label{fig:slopevsE}
\end{figure}

\begin{figure}
\centering
\includegraphics[width=3.2in]{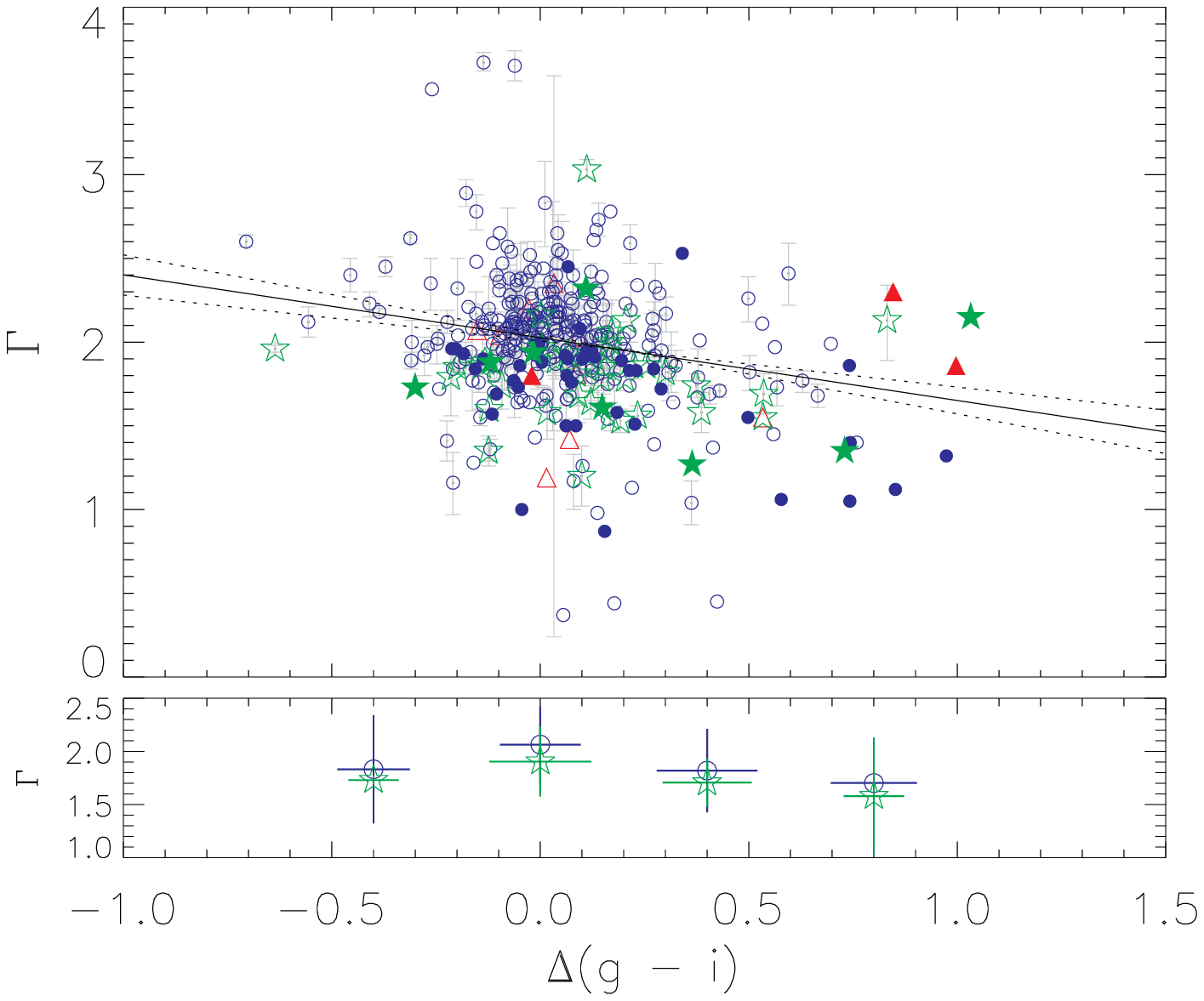}
\includegraphics[width=3.2in]{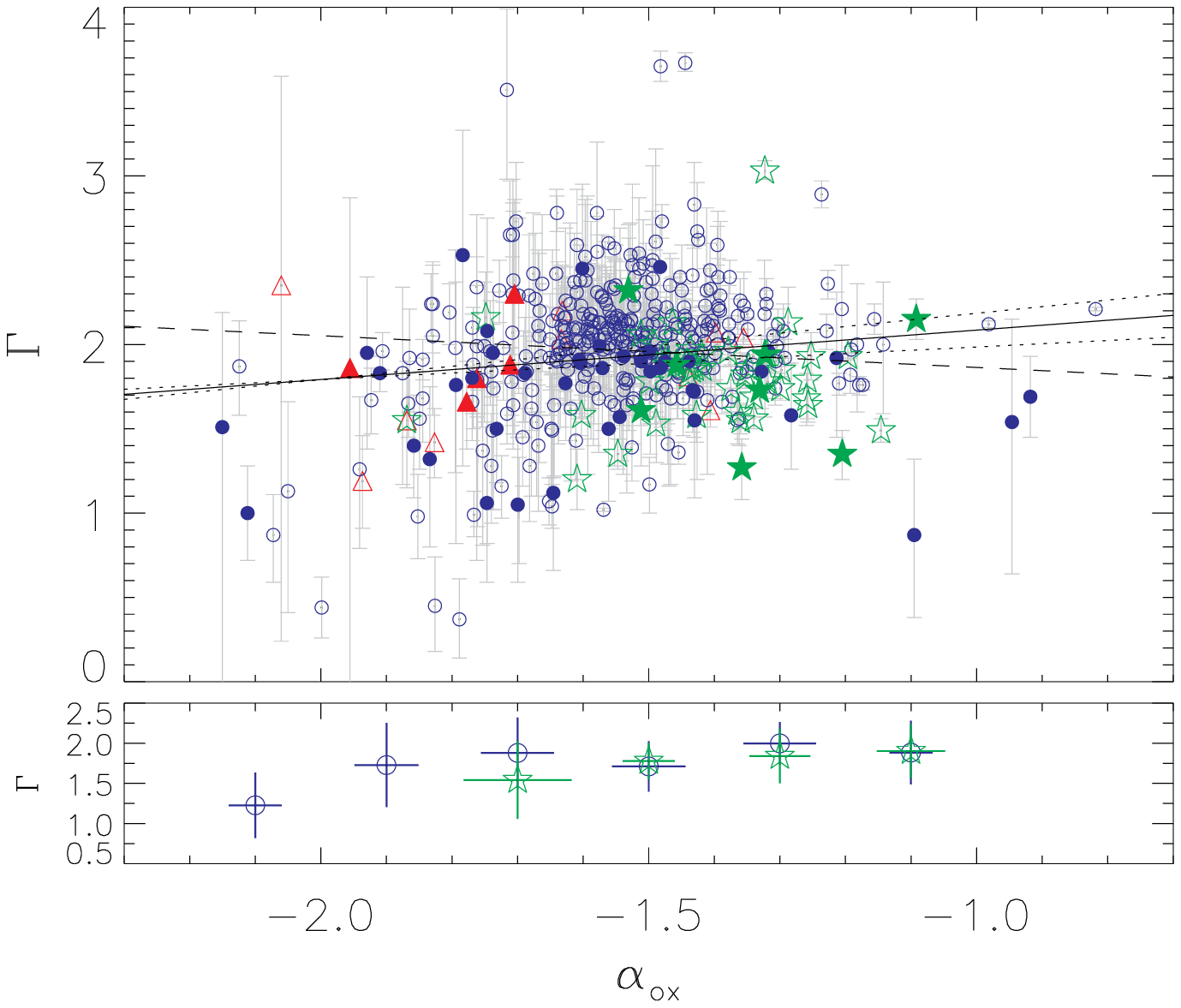}
\includegraphics[width=3.2in]{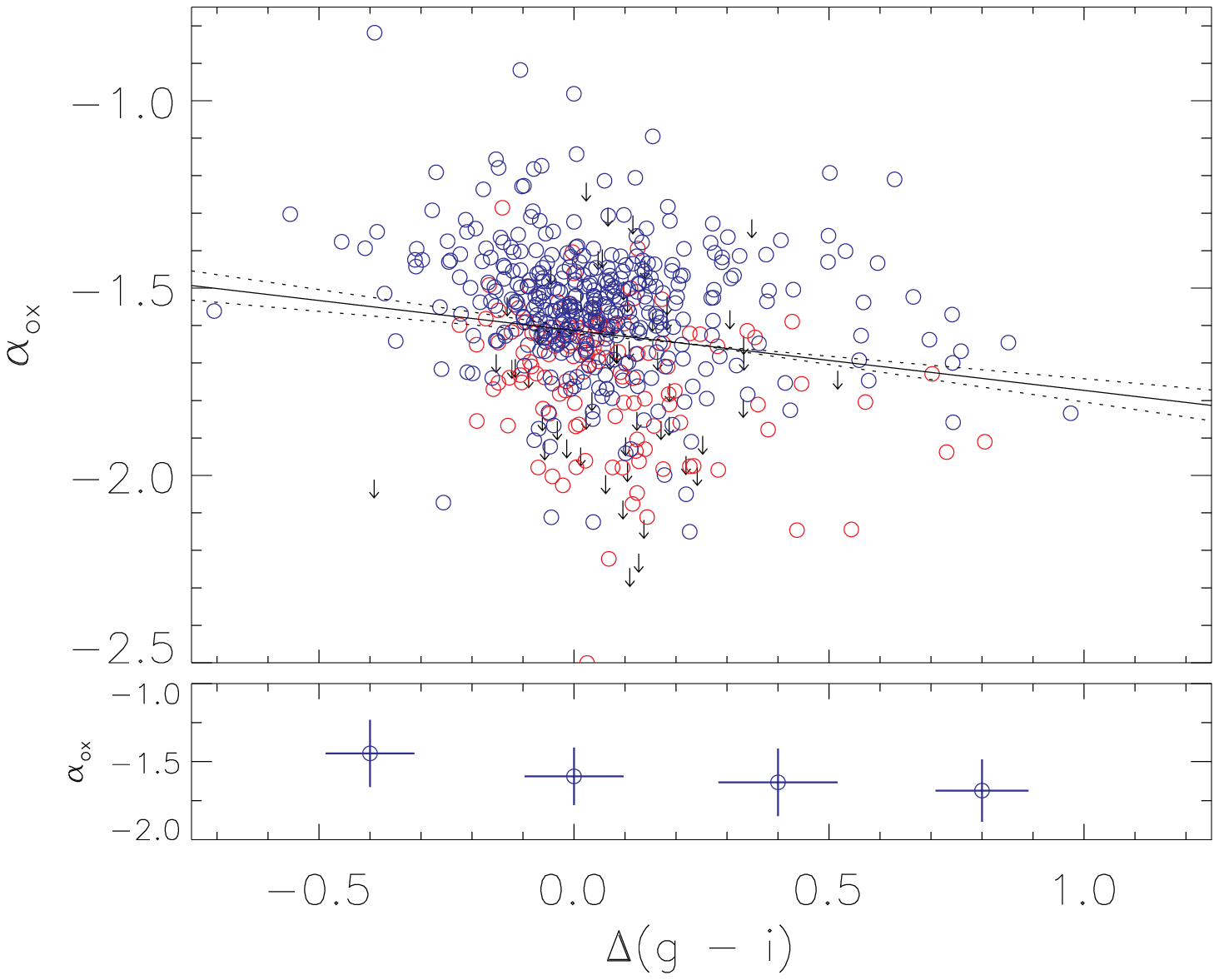}
\caption{TOP:  (a) $\Gamma$ vs. $\alpha_{ox}$, with the $\Gamma - \alpha_{ox}$ correlation found by 
\citet{Green08} plotted for comparison as a dashed line.  (b) $\Gamma$ vs. $\Delta$($g - i$), the relative 
($g - i$) color, for sources with z $<$ 2.3.  Symbols for (a) and (b) are as in Figure \ref{fig:GammaLx}, 
though the correlations are determined only for RQ, non-BAL and non-absorbed quasars.  Weighted least square
regressions are plotted as solid lines, with 3$\sigma$ errors shown as dotted lines.  
(c) $\alpha_{ox}$ vs. relative ($g - i$) color for sources with z $<$ 2.3.  Symbols for (c) are as in Figure 
\ref{fig:aox}.  We plot the OLS regression as a solid line, with 1$\sigma$ errors shown as dotted lines.  
BOTTOM: The weighted mean $\Gamma$ values are given for bin widths as follows: 
(a) $\Delta\alpha_{ox}$ = 0.2, (b) $\Delta$[$\Delta$($g - i$)] = 0.3, and (c) $\Delta$[$\Delta$($g - i$)] = 0.4.  
Error lines mark the 1$\sigma$ dispersions in both axes.  Blue lines represent RQ+non-BAL quasars, while green 
lines represent RL+non-BAL quasars.  Figure (c) does not include BAL or RL quasars.}
\label{fig:Gammacorr}
\end{figure}

\begin{figure}
\centering
\includegraphics[width=3.2in]{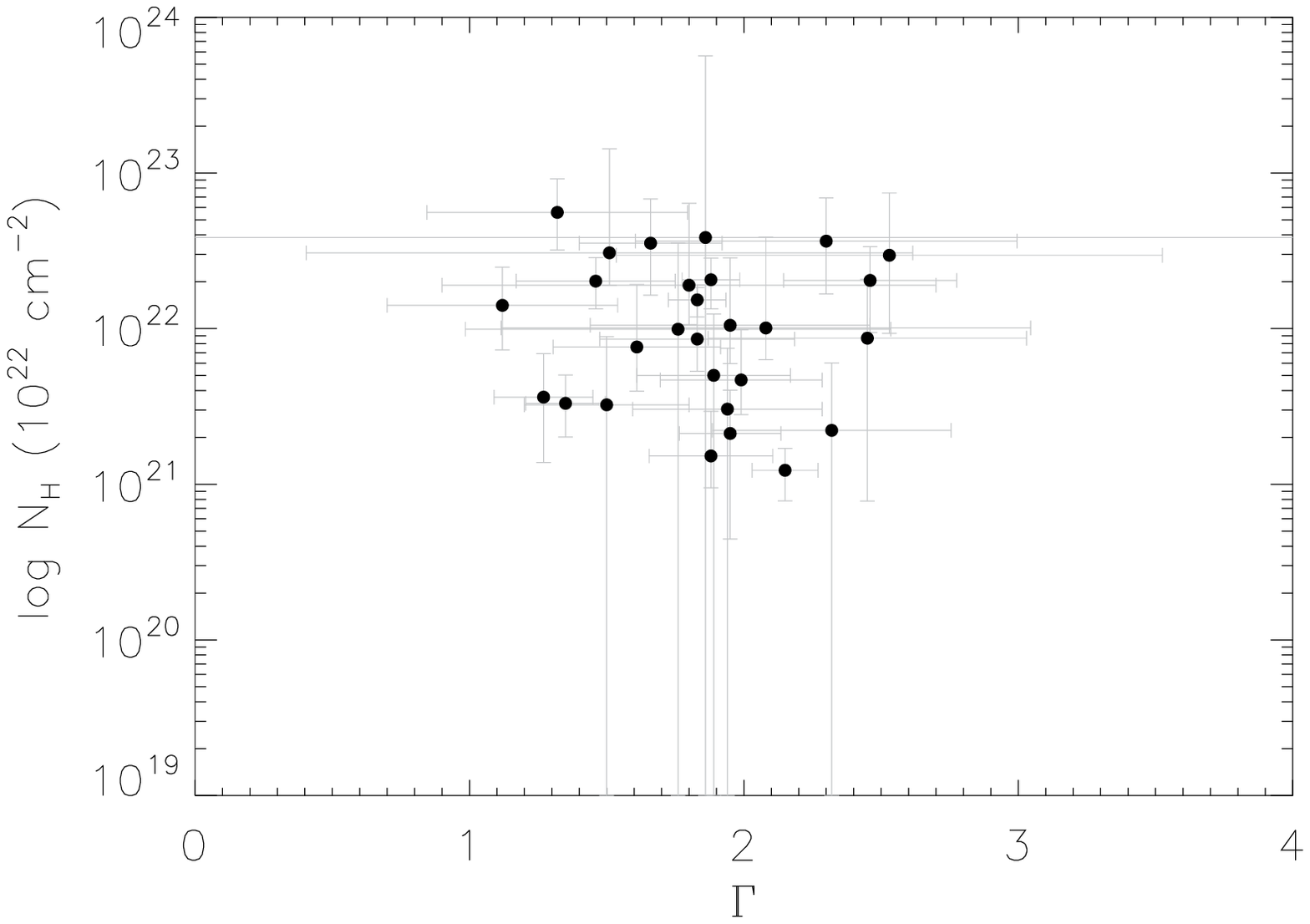}
\includegraphics[width=3.2in]{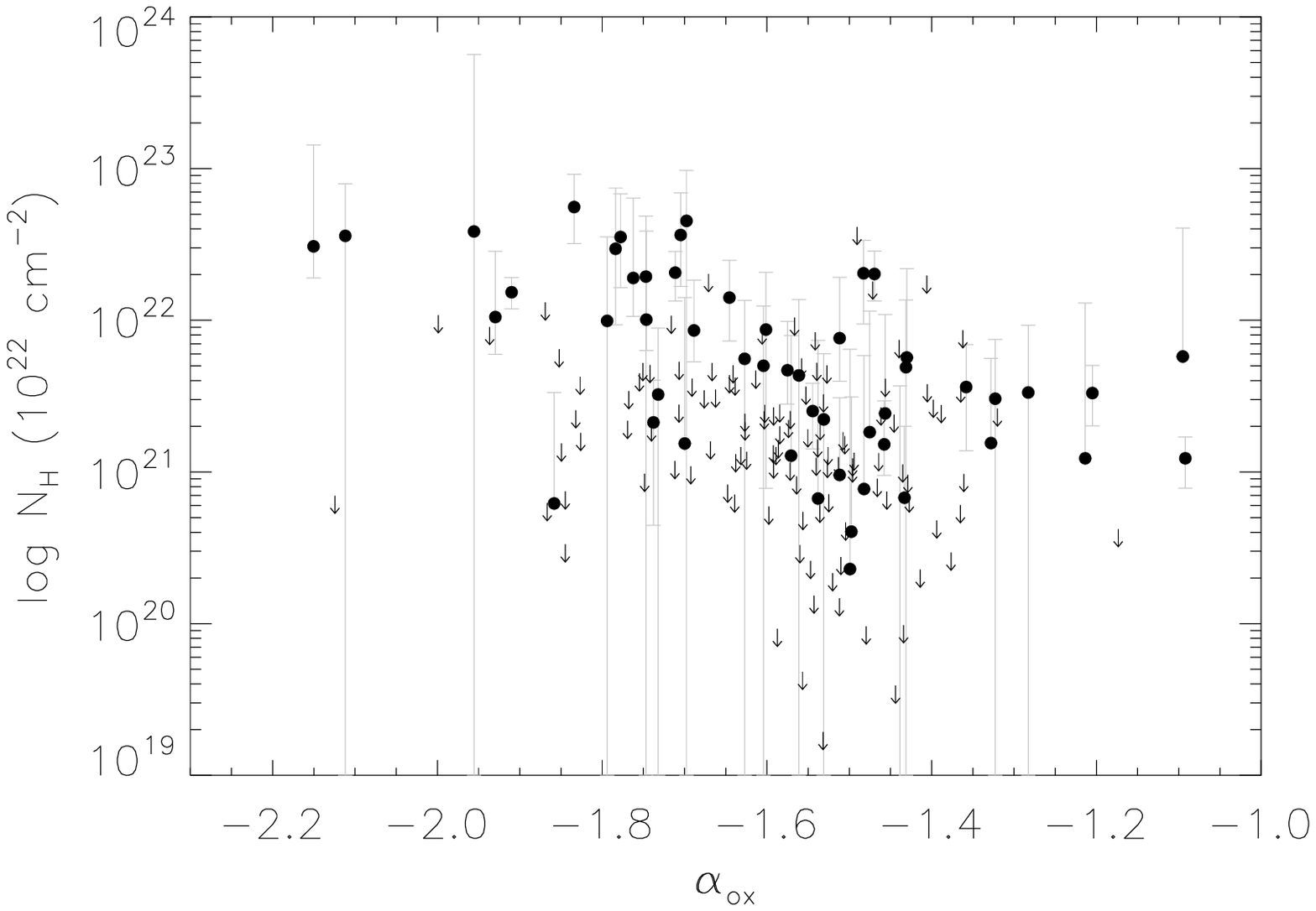}
\includegraphics[width=3.2in]{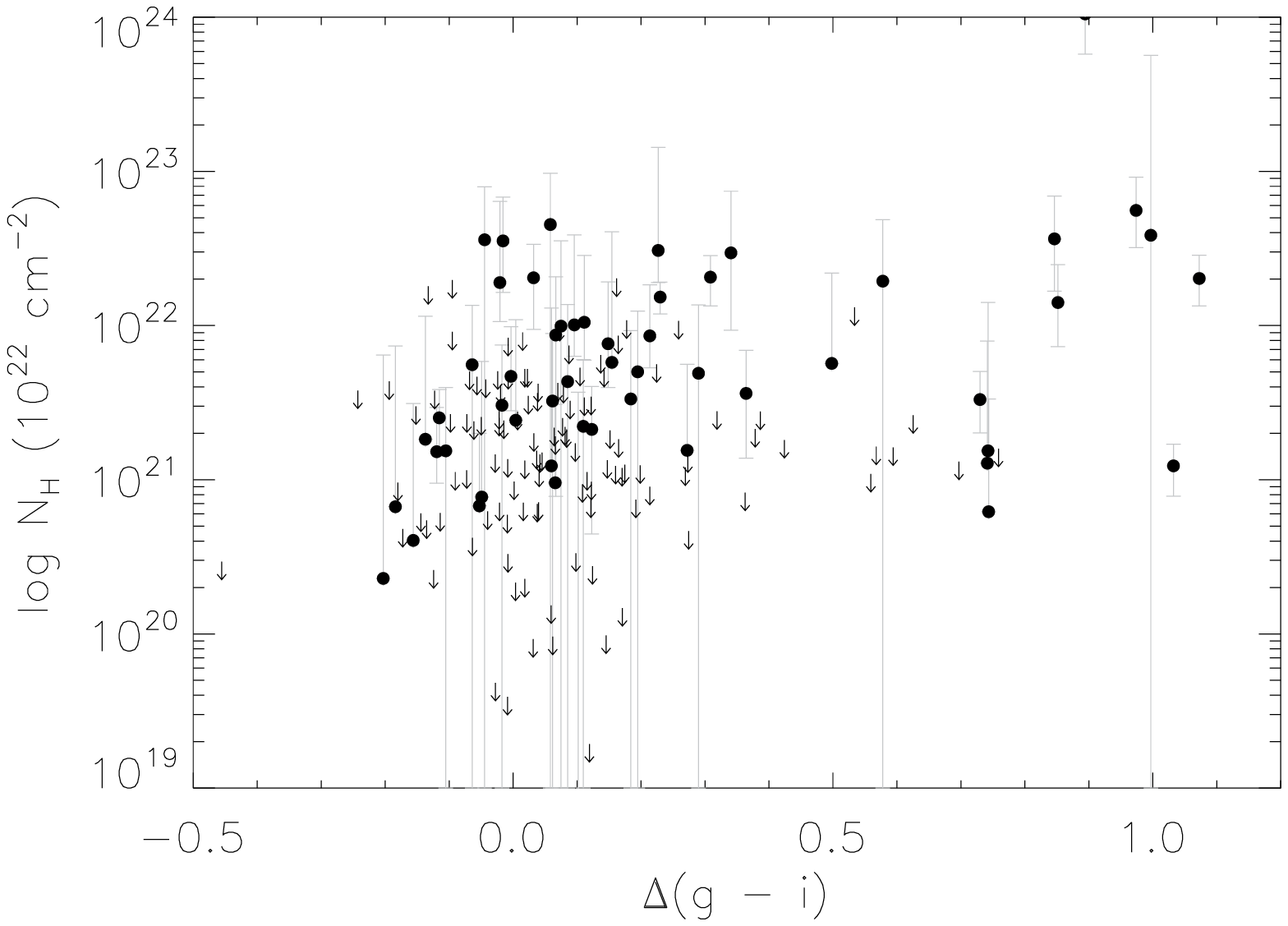}
\caption{(a) Intrinsic absorption (N$_H$) vs. $\Gamma$.  (b) N$_H$ vs. $\alpha_{ox}$.  (c) N$_H$ vs. $\Delta$($g - i$).  
The solid circles in (a) and (b) represent sources preferring the APL model, while in (c) the solid circles 
represent sources preferring either absorption model (FPL or APL).  The arrows represent sources with only a 
90\% upper-limit on N$_H$.  N$_H$ is given in log units of 10$^{22}$ cm$^{-2}$.}
\label{fig:abs}
\end{figure}

\begin{figure}
\centering
\includegraphics[width=3.2in]{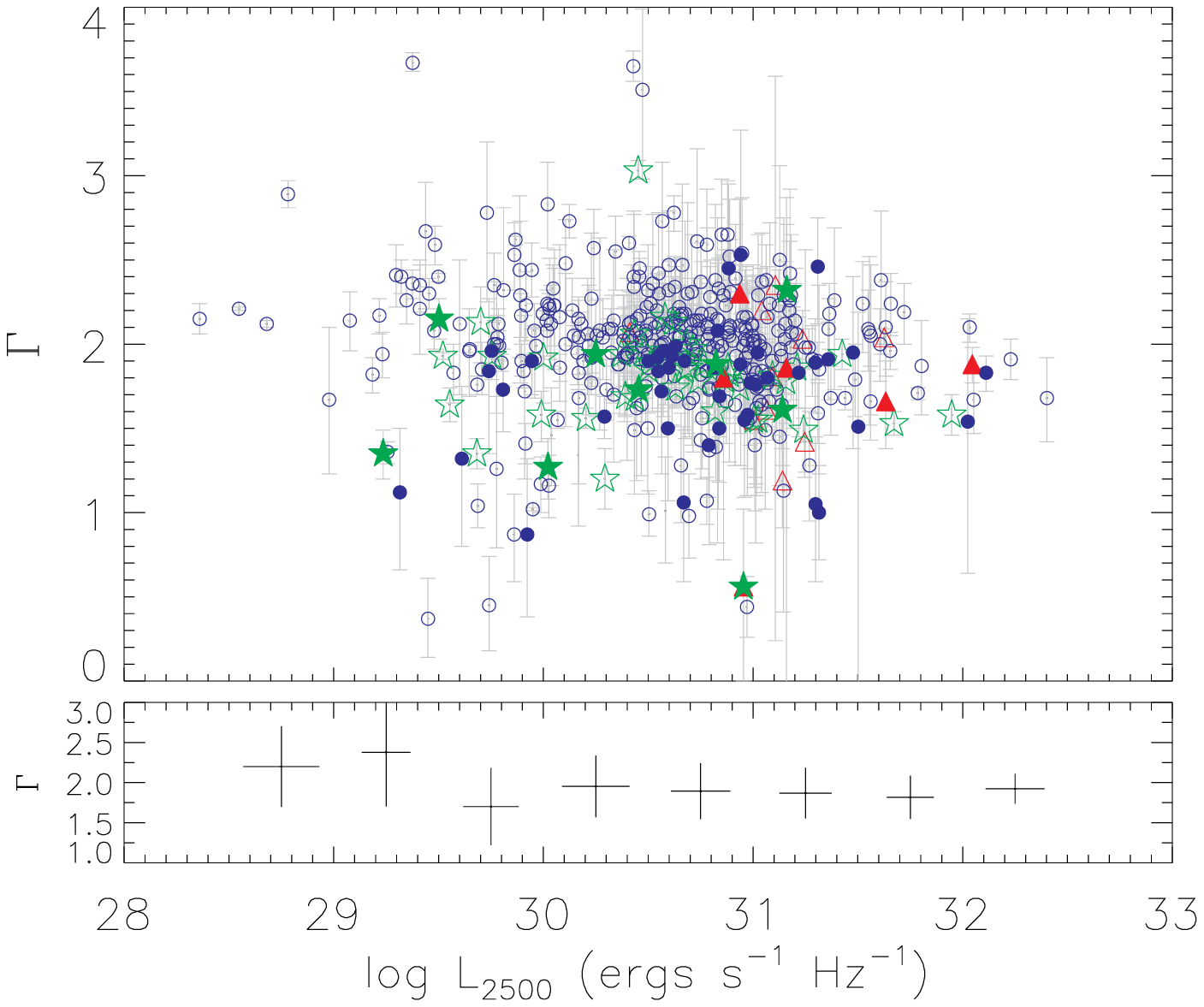}
\includegraphics[width=3.2in]{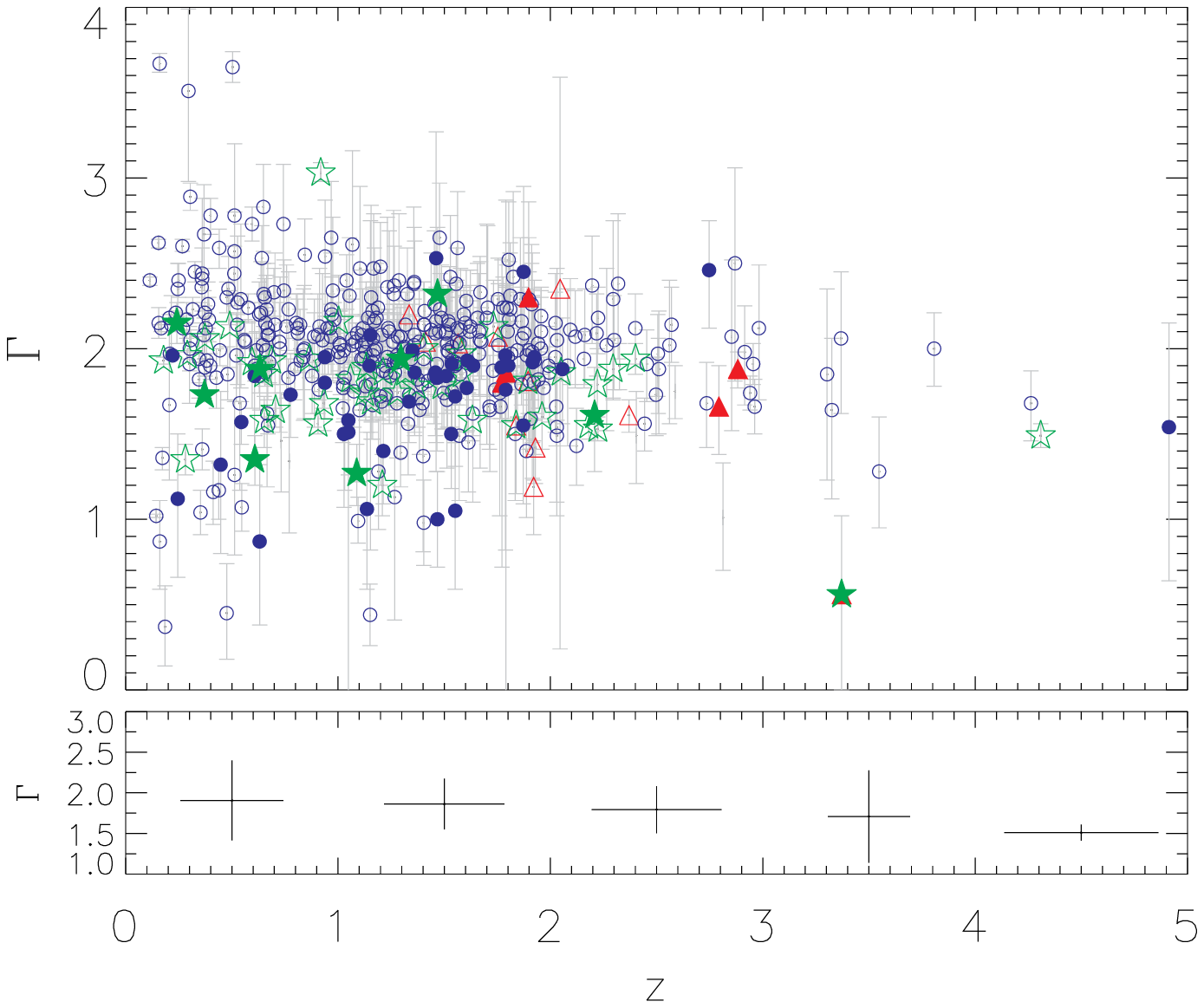}
\caption{X-ray photon index ($\Gamma$) vs. (a) optical 2500 $\mbox{\AA}$ monochromatic luminosity and (b) 
redshift.  Blue circles represent RQ, non-BAL quasars, green stars represent RL quasars, and red triangles 
represent BAL quasars.  Filled symbols are given for those sources that prefer an absorbed power-law with 
an F-test probability P $>$ 0.95.  The bottom plots show the weighted mean $\Gamma$ values for bins of 
width (a) $\Delta$log L$_{2500}$ = 0.5 and (b) $\Delta$z = 1.}
\label{fig:Gammanocorr}
\end{figure}

\end{document}